\documentclass[10pt]{article}
\usepackage{graphicx}

\usepackage{amsfonts}
\usepackage{longtable}
\usepackage[toc,page]{appendix}
\usepackage{amsmath}
\usepackage{amssymb}
\usepackage{eufrak}
\usepackage{epstopdf}
\usepackage{color}

\usepackage[colorlinks, citecolor=blue, linkcolor=blue, backref=true]{hyperref}
\usepackage{longtable}
\usepackage{url}

\usepackage[firstinits=true, sorting=none,parentracker=true, doi=true, isbn=false, natbib=true]{biblatex}

\bibliography{./Bibliography}

\usepackage{upgreek}

\usepackage{amsmath,environ}
\usepackage{mathrsfs}

\newtheorem{remark}{Remark}

\DeclareMathAlphabet{\mathpzc}{OT1}{pzc}{m}{it}

\NewEnviron{eqn}{\begin{equation}\begin{split}
 \BODY
\end{split}\end{equation}
}
\newcommand{\beqans}{\begin{subequations}\begin{eqnarray}}
\newcommand{\eeqans}[1]{\end{eqnarray}\label{#1}\end{subequations}}
\newcommand{\beqan}{\begin{eqnarray}}
\newcommand{\eeqan}{\end{eqnarray}}

\newcommand{\uvec}[1]{\ensuremath{\hat{\mathbf{#1}}}}
\newcommand{\su}{u}
\newcommand{\inc}{i}
\newcommand{\pd}[2]{\frac{\partial#1}{\partial#2}}
\newcommand{\od}[2]{\frac{d#1}{d#2}}
\newcommand{\Iop}{\mathfrak{L}}
\newcommand{\vv}{\mathtt{v}}
\newcommand{\MM}{\mathtt{M}}
\newcommand{\NN}{\mathtt{N}}
\newcommand{\Na}{\aleph}
\newcommand{\Nb}{\NN_b}
\newcommand{\ctheta}{\vartheta}
\newcommand{\sv}{\mathtt{v}}
\DeclareMathOperator{\sL}{\mathpzc{L}}
\newcommand{\la}{\mathrm{b}}
\newcommand*{\bfrac}[2]{\genfrac{}{}{0pt}{}{\raisebox{-.3em}{\scriptsize$#1$}}{\raisebox{.4em}{\scriptsize$#2$}}}
\newcommand{\T}{\mathbb{T}}
\newcommand{\xx}{x}
\newcommand{\yy}{y}
\newcommand{\uu}{u}
\newcommand{\kk}{\upkappa}
\newcommand{\oo}{\upomega}
\newcommand{\varz}{z}
\newcommand{\GG}{\mathcal{G}}
\newcommand{\ZZ}{\mathbb{Z}}
\newcommand{\HH}{\mathpzc{H}}
\newcommand{\AAn}{\mathscr{A}}
\newcommand{\RR}{\mathpzc{R}}
\newcommand{\Ss}{{\mathscr{S}}_0}
\newcommand{\cc}{\mathtt{c}}
\let\Re\relax\DeclareMathOperator{\Re}{\text{\rm Re}}
\let\Im\relax\DeclareMathOperator{\Im}{\text{\rm Im}}

\usepackage[left]{lineno}
\setpagewiselinenumbers

\begin{document}
\title{Wave scattering on lattice structures involving array of cracks}
\author{{\large Gaurav Maurya$^{1}$ and Basant Lal Sharma$^{2}$}\thanks{Department of Mechanical Engineering, Indian Institute of Technology Kanpur, Kanpur, U. P. 208016, India ({bls@iitk.ac.in})}}

\maketitle

\begin{abstract}
{Scattering of waves due to a vertical array of equally-spaced cracks on a square lattice is studied. The convenience of Floquet periodicity reduces the study to that of scattering of specific wave-mode from single crack in a waveguide. The discrete Green's function, for the waveguide, is used to obtain semi-analytical solution for scattering problem in case of finite cracks whereas the limiting case of semi-infinite cracks is tackled by an application of Wiener--Hopf technique. Reflectance and transmittance of such an array of cracks, in terms of incident wave parameters, is analyzed.
Potential applications include construction of tunable atomic scale interfaces to control energy transmission at different frequencies.}
\end{abstract}

\section*{Introduction}
\label{intro}

Multiple scattering \cite{martinbook} has been researched for more than a century and continues to pose interesting {questions},
while simultaneously finding applications 
(see eg., {\cite{julius}}, etc.). In the context of mechanics of solids,
{presence of} defects such as cracks, grooves, holes, etc, \cite{achenbook,julius}, 
lead to scattering of elastic waves.
One of the simplest case occurs for 
scattering of time harmonic anti-plane shear waves as
it often allows an analytical investigation 
\cite{achenbook}; typically, involving two dimensional Helmholtz equation and {the} prescription of Dirichlet or Neumann {condition} on certain boundary.
Same equation also occurs in special situations dealing with acoustic and electromagnetic waves.
Recall that the scattering of H- or E-polarised electromagnetic wave by an infinite array of 
{parallel}
plates 
{was} originally formulated and solved by Carlson and Heins 
{\cite{HeinsI,HeinsII,HeinsIII}}.
In mechanical framework too, such problems have been studied (see \cite{angelI,kentplates,grating,screens}, and references therein), for instance, scattering due to array of cracks. 

In recent years, with advancements in technology, the size of structures has been reduced to a few micrometres or nanometres.
In a simplified setting, {such} structures can be modelled using discrete framework \cite{slepyan,brill}
which has been around for a while \cite{Lifshitz1,
mara};
in fact, some primitive aspects of such models can be traced back to Newton and Hamilton. The discrete models have been extensively used to study brittle fracture \cite{thomson1971lattice,slepyananti,marder,slepyan,marder1}, and recently in a series of articles on {discrete scattering}
\cite{sK, sFK,sC,sFC,sWaveguide} 
in different geometries \cite{sMixed,BGpair1,gmtwocracksasymp}. 
In this {framework},
a {discrete} analogue of the boundary conditions \cite{sK,sC} in the continuous case, depending on the nature of the defects, 
need to be invoked. For example, a crack \cite{sK,sFK} is modelled by assuming broken bonds between two consecutive rows \cite{slepyananti,slepyan}.

The present article follows upon the work of Carlson and Heins \cite{HeinsI,HeinsII,HeinsIII}, in the arena of discrete models, as wave scattering due to finite as well as semi-infinite cracks is investigated; the latter {employs} the method of Wiener and Hopf \cite{noble}.
From the viewpoint of applications, the wave transmission across a periodic arrangement of cracks finds potential relevance in radio frequency devices \cite{yang2015II,yang2015I,yang2016}. 
In certain systems \cite{shin2015}, 
such phononic crystals enable the tailorability, controllability and high conversion efficiency at large frequencies. 
\begin{figure}[h]
\centering
\includegraphics[width=.3\textwidth]{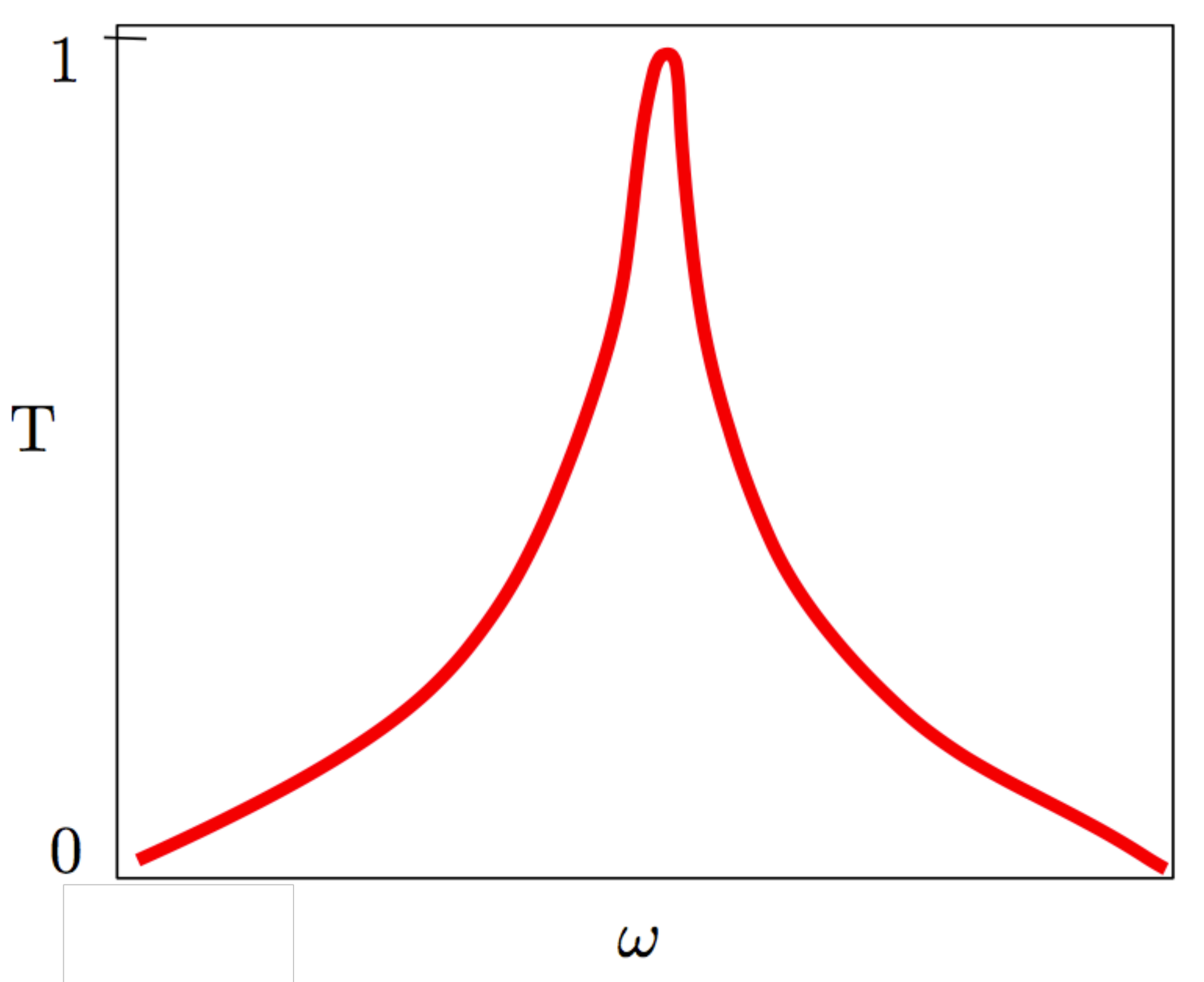}
\caption{
A narrow band of frequency transmitted by the system proposed by 
\cite{shin2015}.}
\label{trans_plotall}
\end{figure}
Although, the recently reported transmission behaviour \cite{shin2015} (particularly, a narrow transmission band, {shown in the schematic of Fig. \ref{trans_plotall}}) is different from the one analyzed in the present work, the geometric arrangements of the cracks allows a favourable transmission/blocking of high frequency lattice waves.
In the context of thermal conduction in nano-structures \cite{anufriev2017,li2015}, the phonon transmission and reflection has been found to be appropriately controlled using a periodic arrangement of discrete scatterers (air holes, typically). 
The present study does not investigate any mechanisms enabling the transduction between photons and phonons or the details of phonon transport in {monolayers} \cite{yang2015I,yang2015II,yang2016,huang2017,xu2018}.

In this article, \S\ref{slm} provides the lattice model. \S\ref{reduce} formulates the scattering due to a single crack on a lattice `waveguide' 
and presents the semi-analytical solution for finite crack; the elementary details of calculation of suitable Green's function are included. The exact solution for semi-infinite case
is given in \S\ref{semiinfinite},
whereas \S\ref{numresult} provides some {key} results and relevant discussion overall. Concluding remarks, and three appendices 
appear at the end of article.

\section{Square lattice model}
\label{slm}

\begin{figure}[hbt!]
\centering
\includegraphics[width=\linewidth]{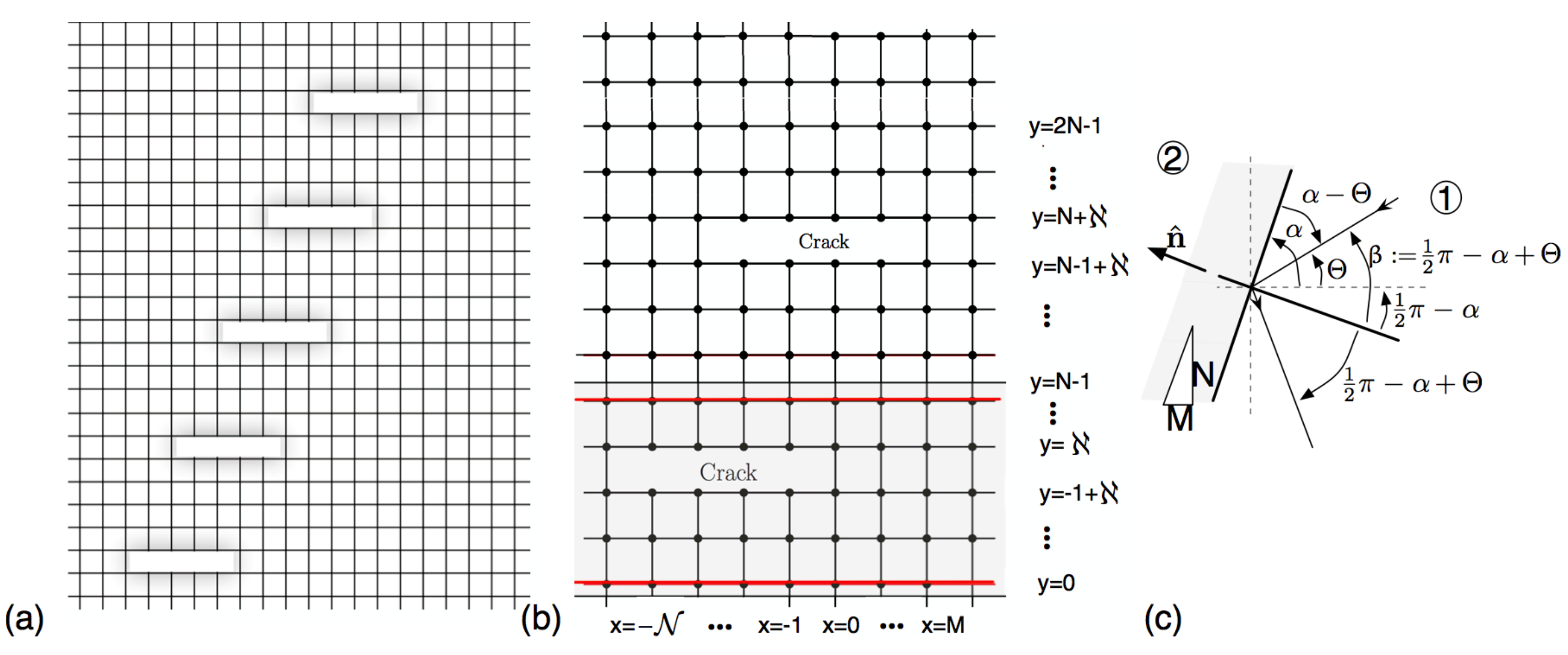}
\caption[Schematic of lattice with array of cracks]{(a) Schematic of lattice with an infinite array of finite, staggered cracks (with $
\Nb=4, \NN=5, \MM=2$). (b) The shaded portion (within red lines)
corresponds to lattice waveguide containing a single crack. (c) {Typical angles}.
}
\label{figone}
\vspace{-5mm}
\end{figure}

Consider an infinite square lattice, with each particle of unit mass and an interaction with its four nearest neighbours through linearly elastic identical, massless bonds with a spring constant $1/b^2$ \cite{sK} (see Fig. \ref{figone}(a)). 
Let $\ZZ$ denote the set of integers, let $\ZZ^2$ denote $\ZZ\times\ZZ$.
The lattice contains an infinite array of finite-length cracks (of length $\Nb$, i.e., the number of broken bonds) described {by} the crack faces
\begin{equation}
\{(\xx,\yy)\in\ZZ^2|-\Nb+n\MM\leq \xx\leq -1+n\MM,\yy=n\NN+\Na\,\text{or}\,\yy=n\NN+\Na-1,n\in\ZZ\},
\label{Sigmak1}
\end{equation}
where $\Nb\in\ZZ$, $\MM\in\ZZ^+$, and $\NN\in\ZZ^+$ with (no loss of generality)
\begin{equation}
\Na=
{\NN/2\text{ when $\NN$ is even whereas } \Na=(\NN-1)/2\text{ when $\NN$ is odd}.}
\label{Nadef}
\end{equation}
Suppose $\uu^i$ describes the incident lattice wave with frequency $\oo$ and a wavenumber $\kk$ which is incident on the lattice at an angle $\Theta\in(-\pi,\pi]$. 
The total displacement $\uu^t$ of a particle satisfies the discrete Helmholtz equation \cite{sK}
\begin{equation}
\Delta \uu^t_{\xx,\yy}+\oo^2\uu^t_{\xx,\yy}=0,
\label{DH}
\end{equation}
away from the crack faces \eqref{Sigmak1}, with $\Delta u_{x,y}=u_{x+1,y}+u_{x-1,y}+u_{x,y+1}+u_{x,y-1}-4u_{x,y}$.
Specifically, it is assumed that {(in terms of the incident angle $\Theta$ and incident wavenumber $\kk$)}
\begin{equation}
\uu^i_{\xx,\yy}={{\mathrm{A}}}~{\exp}({-i\kk \xx \cos\Theta-i\kk \yy\sin\Theta-ib^{-1}\oo t}),~(\xx,\yy)\in\ZZ^2,
\label{uinc}
\end{equation}
where ${{\mathrm{A}}}\in\mathbb{C}$.
Throughout the article, 
$\mathbb{C}$ denotes the set of complex numbers, the real part, $\Re{\varz}$, of a complex number $\varz\in\mathbb{C}$ is denoted by $\varz_1$, and its imaginary part, $\Im{\varz}$, is denoted by $\varz_2$ (so that $\varz=\varz_1+i\varz_2$); $|\varz|$ denotes the modulus for $\varz\in\mathbb{C}$ while $\arg\varz$ denotes the argument for $\varz\in\mathbb{C}$. 
With schematic illustration in Fig. \ref{figone}(c), using the reference to the lattice structure shown in Fig. \ref{figone}(a,b), 
let ${\upalpha}$ denote the angle of stagger of the crack array \eqref{Sigmak1} relative to the ${x}$ axis, i.e.
\begin{eqn}
\tan{\upalpha}={{{\mathtt{N}}}/{{\mathtt{M}}}},\quad
\text{whereas }{\upbeta}{:=}
{\pi/2}-{\upalpha}+{\Theta};
\label{normalincang}
\end{eqn}
here, ${\upbeta}$ is the angle of incidence relative to the outward normal to the `line' of the edges. 
Accordingly, the angle of incidence with respect to the upper side of the edge plane is ${\upalpha}-{\Theta}$.

Substituting \eqref{uinc} in \eqref{DH} (for an intact lattice), a relation for the triplet $\oo$, $\kk$, $\Theta$, called the 
{{\em dispersion relation}} (see \cite{sC}), is obtained;
{it} is given by
$\oo^2=4\sin^2(\frac{1}{2}\kk \cos \Theta)+4\sin^2(\frac{1}{2}\kk \sin \Theta).$
For convenience, a {vanishing} amount of damping is introduced in the model as in \cite{noble}, therefore,
\begin{equation}
\oo=\oo_1+i\oo_2, \oo_2>0.
\label{damping}
\end{equation}
Thus, $\kk$ is also a complex,
$\kk=\kk_1+i\kk_2, \kk_2>0$
(typically, we consider $\oo_2\to0^+$, $\kk_2\to0^+$).
In this article, the scattered wave displacement $\uu$ is defined as the difference between the total displacement $\uu^t$ and the incident wave displacement $\uu^i$ of an arbitrary particle on the lattice:
\begin{equation}
\uu_{\xx,\yy}=\uu_{\xx,\yy}^t-\uu_{\xx,\yy}^i,~~(\xx,\yy)\in\ZZ^2.
\label{ut}
\end{equation}
Following \cite{sK,sFK}, for a particular crack (say, between $\yy=\Na+n\NN$ and $\yy=\Na-1+n\NN$, when $\Na$ is given by \eqref{Nadef} while $n$ is {\em an arbitrary} integer), the force in the vertical bonds connecting the particles at $\yy=\Na+n\NN$ and $\yy=\Na-1+n\NN$, ahead of the crack, is defined by
\begin{eqn}
\hspace{-.12in}
v_\xx^t(n){:=}\frac{-1}{b^2}\vv_\xx^t(n),~\xx\in\ZZ\setminus\{-1+n\MM,\dotsc,-\Nb+n\MM\}, 
\text{with}\, \vv_\xx^t(n){:=}\uu^{{t}}_{\xx,\Na+n\NN}-\uu^{{t}}_{\xx,\Na-1+n\NN}.
\label{force}
\end{eqn}
The force on the particle at $(\xx,\Na+n\NN)$, $\xx\not\in\{-1+n\MM,\dotsc,-\Nb+n\MM\}$, due to the vertical bond with $(\xx,\Na-1+n\NN)$, is $v_x^t(n)$, while that the force at $(\xx,\Na-1+n\NN)$ due to the same bond is $-v_\xx^t(n)$. Since the crack is modelled by assuming broken bonds between two consecutive lattice rows, 
$v_\xx^t(n)=0,~~\xx\in\{-1+n\MM,\dotsc,-\Nb+n\MM\}.$
It is also useful to define the difference of the scattered displacements, $\uu_{\xx,\Na+n\NN}$ and $\uu_{\xx,\Na-1+n\NN}$ as
\begin{equation}
\vv_\xx(n)=\uu_{\xx,\Na+n\NN}-\uu_{\xx,\Na-1+n\NN},~~\xx\in\ZZ.
\label{scatterediff}
\end{equation}
In analogy with \eqref{force}, a part of the force $v_\xx^t(n)$
occurs due to the scattered displacement of particles at $(\xx,\Na+n\NN)$, $\xx\not\in\{-1+n\MM,\dotsc,-\Nb+n\MM\}$; this is given by $v_\xx(n)=-(1/b^2)\vv_\xx(n)$. Let the incident crack opening displacement at $(\xx,\Na)$ and $(\xx,\Na-1)$, 
be denoted by
\begin{equation}
\vv_\xx^i(n)=\uu_{\xx,\Na+n\NN}^i-\uu_{\xx,\Na-1+n\NN}^i.
\label{incdiff}
\end{equation}
Then, $v_\xx^i(n)=-(1/b^2)\vv_\xx^i(n)$ can be interpreted as an `external force' on particle at $(\xx,\Na+n\NN)$, $\xx\in\{-1+n\MM,\dotsc,-\Nb+n\MM\}$. \par
By the virtue of \eqref{DH}, \eqref{uinc} and \eqref{ut} the scattered wave field also satisfies the discrete Helmholtz equation \eqref{DH} (replace $\uu^t$ by $\uu$) 
away from the array of cracks. The displacement field 
on the crack face at $\yy=\Na+n\NN$ and $\yy=\Na-1+n\NN$ satisfies, respectively,
\begin{eqnarray}
\uu_{\xx+1,\Na+n\NN}+\uu_{\xx-1,\Na+n\NN}+\uu_{\xx,\Na+1+n\NN}+(\oo^2-3)\uu_{\xx,\Na+n\NN}=
-\vv_\xx^i(n),
\label{crackbcone}\\
\uu_{\xx+1,\Na-1+n\NN}+\uu_{\xx-1,\Na-1+n\NN}+\uu_{\xx,\Na-2+n\NN}+(\oo^2-3)\uu_{\xx,\Na-1+n\NN}=
\vv_\xx^i(n)
\label{crackbctwo},
\end{eqnarray}
for $\xx\in\{-1+n\MM,\dotsc,-\Nb+n\MM\}$. Here \eqref{crackbcone} and \eqref{crackbctwo} can be interpreted as {\em boundary} conditions for \eqref{DH}. Then, using the definition of the scattered field $\uu$, \eqref{ut}, along with the definitions of $\vv_\xx(n)$ and $\vv_\xx^i(n)$ and the boundary conditions \eqref{crackbcone}, \eqref{crackbctwo}, the 
linear difference
equation \cite{levy} formally
satisfied by the scattered displacement $\uu$ is 
\begin{equation}
\Delta \uu_{\xx,\yy}+\oo^2\uu_{\xx,\yy}=-\sum\nolimits_{n=-\infty}^{\infty}\sum\nolimits_{l=-\Nb}^{-1}(\vv_l(n)+\vv_l^i(n))\delta_{l+n\MM,\xx}(\delta_{\Na+n\NN,\yy}-\delta_{\Na-1+n\NN,\yy}),
\label{crackeq}
\end{equation}
where $\{\vv_l(n)\}_{l=-\Nb, \dotsc, -1; n\in{\ZZ}}$ are an {\em infinite number of unknowns}.
Throughout this article,
the symbol $\delta$ denotes the Kronecker delta so that $\delta_{a, b}$ equals $0$ if $a\ne b$ while it equals $1$ if $a=b$.

\section{Reduction 
to 
lattice waveguide with 
`Floquet boundary': Green's function
and
solution for finite cracks}
\label{reduce}
Since the array of finite-length cracks extends indefinitely, there is a periodicity induced into the system by virtue of the Floquet--Bloch theorem. 
This conveniently reduces the scattering problem to the study of scattering of the incident wave \eqref{uinc} by a single crack in a 
subset $\Ss$ (defined below) with $\NN$ rows (see the shaded region of Fig. \ref{figone}(b)). Suppose the region $\Ss$ corresponds to the crack $n=0$ in \eqref{Sigmak1}. Henceforth, in the context of the symbols used for crack opening displacement, $(n)$ notation is dropped;
$(0)$ will be omitted for making reference to $\Ss$.
Thus, a shorter notation and Floquet--Bloch periodicity based reduction allows a simplification from
the system of equations \eqref{crackeq}
to the following equation
\begin{equation}
\Delta \uu_{\xx,\yy}+\oo^2\uu_{\xx,\yy}=-\sum\nolimits_{l=-\Nb}^{-1}(\vv_l+\vv_l^i)\delta_{l,\xx}(\delta_{\Na,\yy}-\delta_{\Na-1,\yy}), (\xx,\yy)\in\Ss.
\label{crackeqn}
\end{equation}
Observe that the set $\Ss$ of lattice sites is infinite in the horizontal direction while it is confined in the vertical direction. 
We employ the natural notation ${\ZZ}_{a}^{b}$ for the set $\{a, a+1, \dotsc, b\}$ ($\subset{\ZZ}$).
Indeed, 
$\cup_{n\in{\ZZ}}{\mathscr{S}}_n={\ZZ^2}\text{ with }{\mathscr{S}}_n={\mathscr{S}}_0+n{\mathtt{N}}\uvec{j}=\{({x}, {y}+n{\mathtt{N}})\in{{\ZZ^2}}: {x}\in{\ZZ}, {y}\in{\ZZ}_0^{{\mathtt{N}}-1}\}.$

The incident lattice wave {\eqref{uinc}, i.e., $\uu_{\xx,\yy}^i={{\mathrm{A}}}\exp({-i\kk(\xx\cos\Theta+\yy\sin\Theta)})$} in ${\mathscr{S}}_n$, i.e., at one set of $\NN$ rows, in the lattice is related to another set ${\mathscr{S}}_{n+1}$ via
\begin{eqnarray}
\uu_{\xx+\MM,\yy+\NN}^i={{\uppsi}}\uu_{\xx,\yy}^i,
\text{where }
{{\uppsi}}={\exp}({-i\kk(\MM\cos\Theta+\NN\sin\Theta)}).
\label{floquet}
\end{eqnarray}
By the Floquet--Bloch theorem, the scattered wave field must satisfy identical condition
\begin{eqn}
{\su}_{{x}+{\mathtt{M}}, {y}+{\mathtt{N}}}&={\uppsi}{\su}_{{x}, {y}}.
\label{periodconditioncrackgen}
\end{eqn}
In the perspective of the infinite square lattice, the formal definition of $\Ss$ is 
\begin{equation}
\{(\xx, \yy)\in\ZZ^2| \yy
\in{\ZZ}_0^{{\mathtt{N}}-1}, \uu_{\xx+\MM,\yy+\NN}={{\uppsi}}\uu_{\xx,\yy}\}.
\label{waveguide}
\end{equation}
Notably $\Ss$ includes the `Floquet' periodic boundary conditions inherently.
The periodically repeating cell (as ${\mathscr{S}}_n$s are {\em copies} of ${\mathscr{S}}_0$) is the `waveguide' mentioned earlier. \par
Classically, the wave field in a scattering problem can be written in terms of an appropriate Green's function (see for example, \cite{achenbook}). It has been shown that using discrete Fourier transforms \cite{sFK} a discrete Green's function (following the traditional terminology), can be also used for the lattice wave scattering. In the present case, the discrete Green's function $\GG$ is sought for the lattice waveguide $\Ss$ and it satisfies a difference equation given by
\begin{equation}
\Delta \GG_{\xx,\yy}+\oo^2\GG_{\xx,\yy} = \delta_{\xx_0,\xx}\delta_{\yy_0,\yy}, ~~(\xx,\yy)\in \Ss,
\label{DHG}
\end{equation}
where it is assumed that a source is located at $(\xx_0,\yy_0)\in\Ss$. Due to \eqref{damping}, note that the Green's function 
$\GG_{\xx,\yy}\sim {\exp}({-\kk_2 |\xx|})$ as $|\xx|\to\infty$. 
The Green's function, the subject of the following, must satisfy the Floquet periodic boundary conditions of the waveguide.
Thus, the difference equation \eqref{DHG} is subjected to the condition
(using \eqref{floquet} and \eqref{periodconditioncrackgen})
$\GG_{\xx+\MM,\yy+\NN}={{\uppsi}}\GG_{\xx,\yy}, (\xx,\yy)\in\Ss.$
For the particles at the boundary rows, i.e., at $\yy=0$ and $\yy=\NN-1$, 
this leads to
$\GG_{\xx+\MM,\NN} = {{\uppsi}}\GG_{\xx,0}, \xx\in\ZZ,$
and
$\GG_{\xx+\MM,\NN-1}={{\uppsi}}\GG_{\xx,-1}, \xx\in\ZZ,$
respectively. Using 
\eqref{DHG}, the governing equation for a particle at the boundary of the waveguide (that is, $\yy=0$ and $\yy=\NN-1$, respectively) can be written as
\begin{eqnarray}
\GG_{\xx+1,0}+\GG_{\xx-1,0}+\GG_{\xx,1}+{{\uppsi}}^{-1}\GG_{\xx+\MM,\NN-1}+(\oo^2-4)\GG_{\xx,0}=0,~\xx\in\ZZ,
\label{bcone}\\
\text{and }
\GG_{\xx+1,\NN-1}+\GG_{\xx-1,\NN-1}+\GG_{\xx,\NN-2}+{{\uppsi}}\GG_{\xx-\MM,0}+(\oo^2-4)\GG_{\xx,\NN-1}=0,~\xx\in\ZZ.
\label{bctwo}
\end{eqnarray}
Suppose that the discrete Fourier transform of a sequence $\{u_m\}_{m\in\ZZ}$ is denoted by $u^{{\mathrm{F}}}$ and defined by $u^{{\mathrm{F}}}(\varz)
=\sum\nolimits_{m=-\infty}^{+\infty}u_m\varz^{-m}$.
Using the discrete Fourier transform (see also \cite{sFK,gmtwocracksasymp}), the transformed Green's function can be written as
(suppressing $\varz$ dependence for brevity)
\begin{equation}
\GG_\yy^{{\mathrm{F}}}=\sum\nolimits_{\xx=-\infty}^{\infty}\GG_{\xx,\yy}\varz^{-\xx}. 
\label{dft}
\end{equation}
Based on the nature of 
$\GG_{\xx,\yy}$ as $|\xx|\to\infty$,
the region of analyticity of above Fourier transform \cite{sWaveguide} can be found to be an annulus ${{\AAn}}_g$ in the complex plane centred at the origin, which is given by 
${{\AAn}}_g=\{\varz\in\mathbb{C}:{\exp}({-\kk_2})<|\varz|<{\exp~}{\kk_2}\}.$
The application of the discrete Fourier transform \eqref{dft} to \eqref{DHG} results into 
\begin{eqnarray}
\GG_\yy^{{\mathrm{F}}}({\HH}+2)-(\GG_{\yy+1}^{{\mathrm{F}}}+\GG_{\yy-1}^{{\mathrm{F}}})=-\varz^{-\xx_0}\delta_{\yy,\yy_0}, \label{Fgov}
\text{where }
{\HH}=2-\varz-\varz^{-1}-\oo^2.\label{HHdef}
\end{eqnarray}
Similarly, the application of the discrete Fourier transform \eqref{dft} to the boundary conditions, \eqref{bcone} and \eqref{bctwo}, respectively, yields 
\begin{equation}
\GG_0^{{\mathrm{F}}}
({\HH}
+2)-(\GG_1^{{\mathrm{F}}}
+{{\uppsi}}^{-1}\varz^{\MM}\GG_{\NN-1}^{{\mathrm{F}}}
)=0, 
\text{and }
\GG_{\NN-1}^{{\mathrm{F}}}
({\HH}
+2)-(\GG_{\NN-2}^{{\mathrm{F}}}
+{{\uppsi}}\varz^{-\MM}\GG_0^{{\mathrm{F}}}
)=0. 
\label{fbconetwo}
\end{equation}
Since \eqref{Fgov} is a non-homogeneous linear difference equation 
in $\yy$ with coefficients independent of $\yy$, the solution 
can be written as \cite{levy}
$\GG_\yy^{{\mathrm{F}}}
=\GG_\yy^{Fh}
+\GG_\yy^{{{\mathrm{F}}}nh},$ where $\GG_\yy^{{{\mathrm{F}}}h}
$ is the solution to the homogeneous equation 
$\GG_\yy^{{{\mathrm{F}}}h}
({\HH}+2)-(\GG_{\yy+1}^{{{\mathrm{F}}}h}+\GG_{\yy-1}^{{{\mathrm{F}}}h})=0,$ 
with certain boundary conditions (to be stated below), and $\GG_\yy^{{{\mathrm{F}}}nh}
$ is a particular solution of 
\begin{equation}
\GG_\yy^{{{\mathrm{F}}}nh}
({\HH}+2)-(\GG_{\yy+1}^{{{\mathrm{F}}}nh}+\GG_{\yy-1}^{{{\mathrm{F}}}nh})=-\delta_{\yy,\yy_0}\varz^{-\xx_0}. 
\label{nonhomo}
\end{equation}
Using elementary calculus \cite{sCheby}, a (particular) solution of \eqref{nonhomo} is found 
\begin{equation}
\GG_y^{{{\mathrm{F}}}nh}
=G_0(\varz){{\lambda}}^{|\yy-\yy_0|}(\varz), \varz\in{{\AAn}},
\label{guess}
\end{equation}
\begin{eqn}
\text{where 
\cite{sK,sFK,slepyan} }
{{\lambda}}({{z}}){:=}\frac{{{\mathpzc{r}}}({{z}})-{{\mathpzc{h}}}({{z}})}{{{\mathpzc{r}}}({{z}})+{{\mathpzc{h}}}({{z}})}, {{z}}\in{\mathbb{C}}\setminus{\mathscr{B}}, 
{{\mathpzc{h}}}({{z}}){:=}\sqrt{{{{\HH}}}({{z}})}, 
{{\mathpzc{r}}}({{z}}){:=}\sqrt{{{{\HH}}}({{z}})+4}.
\label{lambda}
\end{eqn}
The square root function, $\sqrt{\cdot}$, has the branch cut from $-\infty$ to $0$. 
${\mathscr{B}}$ denotes the union of branch cuts for ${{\lambda}}$, borne out of the chosen branch
for ${{\mathpzc{h}}}$ and ${{\mathpzc{r}}}$ such that 
$|{{\lambda}}({{z}})|\le1, 
{{z}}\in{\mathbb{C}}\setminus{\mathscr{B}}.$
In above equations, the annulus 
${{\AAn}}$ is given by 
\begin{equation}
{{\AAn}}={{\AAn}}_g\cap{{\AAn}}_{\Iop},
\label{annulusA}
\end{equation}
with ${{\AAn}}_{\Iop}$ being the annular region where ${\mathpzc{h}}$ and ${\mathpzc{r}}$ (and ${\lambda}$ too) are analytic {(for ${\AAn}_g$, see the sentence following \eqref{dft})}. 
The coefficient $G_0$ in \eqref{guess} is determined by substituting the ansatz of $\GG_y^{{{\mathrm{F}}}nh}$ in \eqref{nonhomo}, that is,
$G_0
{{\lambda}}^{|\yy-\yy_0|}({\HH}+2)-(G_0{{\lambda}}^{|\yy-\yy_0+1|}+G_0{{\lambda}}^{|\yy-\yy_0-1|})=-\varz^{-\xx_0}\delta_{\yy-\yy_0,0},$
for $\varz\in{{\AAn}}$, which leads to $G_0=-{\varz^{-\xx_0}}/{({{\HH}+2-2{{\lambda}}})}$,
so that, a particular solution of the linear non-homogeneous difference equation \eqref{nonhomo} can be written as \cite{levy}
\begin{equation}
\GG_\yy^{{{\mathrm{F}}}nh}=-\frac{\varz^{-\xx_0}{{\lambda}}^{|\yy-\yy_0|}}{{\HH}+2-2{{\lambda}}}, \varz\in{{\AAn}}.
\label{fnh}
\end{equation}
After substitution of the particular solution \eqref{fnh} of \eqref{Fgov} in the boundary conditions \eqref{fbconetwo} (for $\yy=0$ and $\yy=\NN-1$), we obtain the boundary conditions for the homogenous solution $\GG_\yy^{{{\mathrm{F}}}h}$,
\begin{eqnarray}
\GG_{0}^{{{\mathrm{F}}}h}({\HH}+2)-(\GG_1^{{{\mathrm{F}}}h}+ \varz^{\MM}{{\uppsi}}^{-1}\GG_{\NN-1}^{{{\mathrm{F}}}h})
=\frac{\varz^{-\xx_0}({{\lambda}}^{|\yy_0|}({\HH}+2)-{{\lambda}}^{|1-\yy_0|}-{{\lambda}}^{|\NN-1-\yy_0|}{{\uppsi}}^{-1}\varz^{\MM})}{{\HH}+2-2{{\lambda}}},
\label{aA}\\
\GG_{\NN-1}^{{{\mathrm{F}}}h}({\HH}+2)-(\GG_{\NN-2}^{{{\mathrm{F}}}h}+{{\uppsi}}\varz^{-\MM}\GG_{0}^{{{\mathrm{F}}}h})
=\frac{\varz^{-\xx_0}({{\lambda}}^{|\NN-1-\yy_0|}({\HH}+2)-{{\lambda}}^{|\NN-2-\yy_0|}-{{\uppsi}}\varz^{-\MM}{{\lambda}}^{|\yy_0|})}{{\HH}+2-2{{\lambda}}},
\label{bB}
\end{eqnarray}
for $\varz\in{{\AAn}}$ {(recall that $\HH$ is defined by \eqref{HHdef})}; these conditions are used to determine the unknown coefficients in the general solution for homogenous part. 
After an elementary calculation,
we find 
\begin{eqnarray}
{\GG}_{\yy}^{{\mathrm{F}}}
=-\varz^{-\xx_0}\frac{{{\mathtt{U}}}_{\NN-|\yy-\yy_0|-1}(\ctheta)+{{\mathtt{U}}}_{|\yy_0-\yy|-1}(\ctheta)({\uppsi}\varz^{-\MM})^{\text{sign}(\yy-\yy_0)}}{2{\mathtt{T}}_\NN(\ctheta)-({\uppsi}\varz^{-\MM}+{\uppsi}^{-1}\varz^{\MM})},
\label{Gyfinal}
\text{where }
\ctheta=1+\frac{1}{2}{\HH}.
\label{defctheta}
\end{eqnarray}
It is emphasized that the numerator and denominator in \eqref{Gyfinal} {involve} the Chebyshev polynomials which have {been} found significant in the description of wave propagation characteristics of lattice waveguides \cite{sCheby,sHoney}. {In particular, ${\mathtt{U}}_n$ denotes the Chebyshev polynomials of second kind \cite{mason} defined by ${\mathtt{U}}_n(\ctheta):=\sin((n+1)\ctheta)/\sin\ctheta, n\ge0$ while ${\mathtt{T}}_n$ denotes the Chebyshev polynomials of first kind \cite{mason} defined by ${\mathtt{T}}_n(\ctheta):=
\cos n\ctheta$.}

The discrete Green's function in the physical domain can be obtained by inverse Fourier transform {of} 
\eqref{Gyfinal}, i.e.,
\begin{eqnarray}
\GG_{\xx,\yy}=\frac{1}{2\pi i}\oint_{C}\GG_\yy^{{\mathrm{F}}}(\varz)\varz^{\xx-1}\,d\varz,
\label{fourierinvG}
\end{eqnarray}
where $C$ can be chosen to be a closed contour that lies inside the annulus ${{\AAn}}$ \eqref{annulusA}.
Due to the vanishingly small imaginary part of the frequency and hence, the wavenumber, all the singularities of the integrand in \eqref{fourierinvG}, are either inside the unit circle or outside the circle, that is, they are away from the contour (see \cite{noble}). 
Till this point, our exposition completes the derivation of the discrete Green's function. 
It turns out that the denominator of the transformed Green's function, \eqref{Gyfinal}, represents the dispersion relation for a square lattice waveguide with Floquet--Bloch periodic boundaries. 
{Brief} discussion of the same {is} provided in Appendix \ref{wdisperse}.\par 
The description of the scattered displacement field in terms of the Green's function \eqref{fourierinvG} now follows the well known approach \cite{Lifshitz1} (see also \cite{sFK,sFC} for notation relevant to the manipulations presented below).
In fact, the present problem is closely related to the scattering due to a finite crack in infinite lattice and the description of the scattered field $\uu$ in terms of the Green's function has been discussed systematically in \cite{sFK}. 
For additional clarity, $\GG_{\xx,\yy;\xx_0,\yy_0}$ will be used instead of $\GG_{\xx,\yy}$ in the subsequent paragraphs.
Note that the expression \eqref{fourierinvG} is the solution of the equation \eqref{DHG} {which has} the source 
located at $(\xx_0,\yy_0)$ {(recall \eqref{DHG})}.

Using \eqref{crackeqn} and \eqref{DHG}, by inspection, it can be found that the scattered displacement field due to an infinite array of cracks in the lattice is given by
\begin{equation}
\uu_{\xx,\yy}=-\sum\nolimits_{l=-\Nb}^{-1}(\vv_l+\vv_l^i)(\GG_{\xx,\yy;l,\Na}-\GG_{\xx,\yy;l,\Na-1}),
\label{uk}
\end{equation}
for $(\xx,\yy)\in\Ss$. 
In fact, \eqref{uk} provides the unique solution to the equation \eqref{crackeqn} in terms of the crack opening displacement $\{\vv_l\}_{l\in\in{\ZZ}_{-\Nb}^{-1}}$. The rigorous aspects of the issue of existence and uniqueness of the solution, for the assumed case $\oo_2>0$, are analogous to the results provided in \cite{sFK} and are, therefore, omitted in the present article. 
Substitution of \eqref{uk} into \eqref{scatterediff} yields a system of equations of the form
\begin{eqnarray}
\sum\nolimits_{l=-\Nb}^{-1}\cc_{j,l}\vv_l=b_j,~~j\in{\ZZ}_{-\Nb}^{-1}, 
\label{systemK}\\
\text{where }
\cc_{j,l}=\delta_{j,l}-(-\GG_{j,\Na;l,\Na}+\GG_{j,\Na;l,\Na-1}+\GG_{j,\Na-1;l,\Na}-\GG_{j,\Na-1;l,\Na-1}),\\
b_j=\sum\nolimits_{l=-\Nb}^{-1}(\delta_{l,j}-\cc_{j,l}),\text{ for }j\in{\ZZ}_{-\Nb}^{-1}. 
\end{eqnarray}
Introducing 
$\mathbf{v}=[\vv_{-1},\vv_{-2},\dotsc,\vv_{-\Nb}]^T\in\mathbb{C}^{\Nb}~~\text{and}~~\mathbf{v}^i=[\vv_{-1}^i,\vv_{-2}^i,\dotsc,\vv_{-\Nb}^i]^T\in\mathbb{C}^{\Nb},$
\eqref{systemK} can be expressed as $\mathbf{v}=\mathbf{F}_{\Nb}(\mathbf{v}+\mathbf{v}^i)$, where $\mathbf{F}_{\Nb}$ is an $\Nb\times \Nb$ matrix with $[\mathbf{F}_{\Nb}]_{j,l}=
{\delta_{j,l}-\cc_{j,l}}$. Therefore,
$\mathbf{v}=(\mathbf{I}_{\Nb}-\mathbf{F}_{\Nb})^{-1}\mathbf{F}_{\Nb}\mathbf{v}^{i}.$
The matrix $\mathbf{F}_{\Nb}$ is a matrix, of the Toeplitz form \cite{sK,sFK} due to peculiar nature of the Green's function \eqref{DHG}. Eventually, the complete displacement field on the lattice (with an array of cracks) 
can be written by substitution of the components of $\mathbf{v}$
{in}
\eqref{uk} and extending to the entire lattice by using the Floquet phase factor \eqref{floquet}.

\section{Semi-infinite cracks: Wiener--Hopf method}
\label{semiinfinite}

\begin{figure}[h]
\centering
\includegraphics[width=\textwidth]{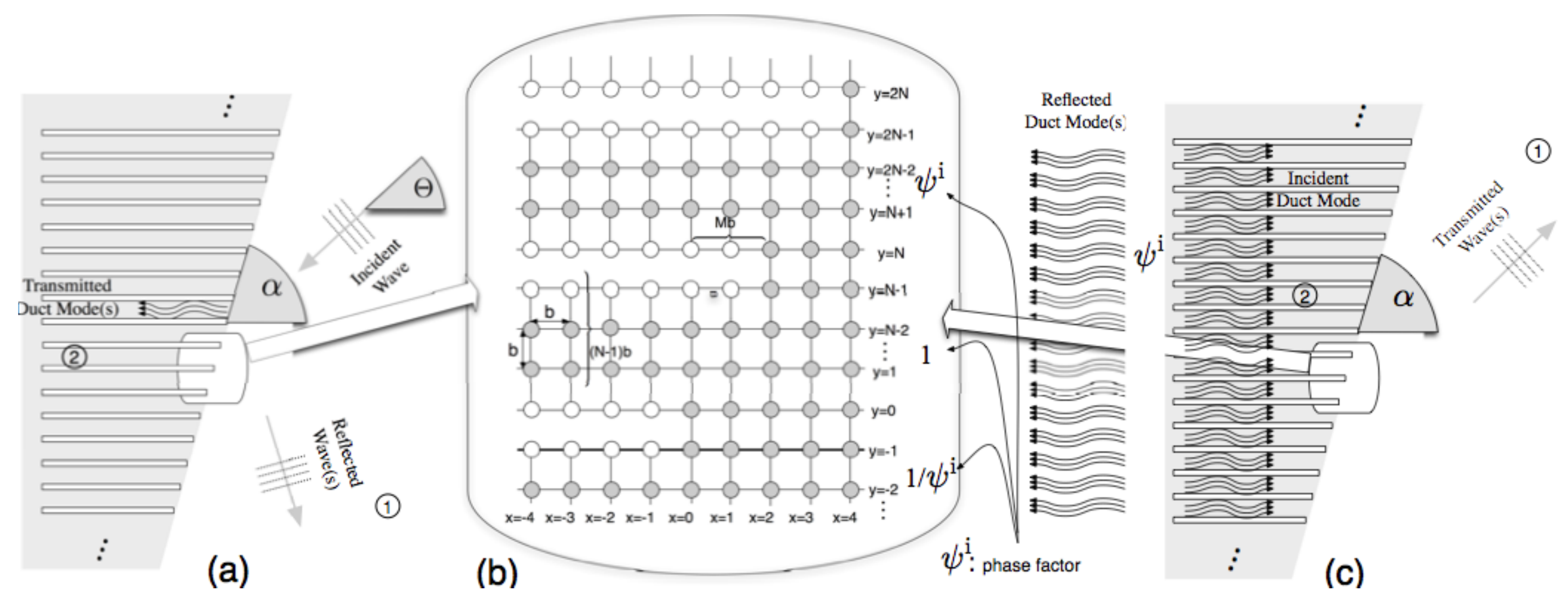}
\caption{A schematic illustration of an infinite array of cracks with a horizontal stagger ${\mathtt{M}}$  (b), together with directions of incident wave, reflected wave(s), and duct mode(s) in (a) and (c), respective to the incidence from bulk lattice (1) and portion between the cracks (2). 
The crack tips shown schematically in (b). Intact lattice is shown as gray dots and the particles located at the crack faces
(lacking a nearest neighbor bond) as white dots. }
\label{squarelatticearraycracks}
\end{figure}

In the following, the limiting case as $\Nb\to\infty$, i.e. semi-infinite cracks, is analyzed via the method of Wiener and Hopf \cite{noble}. We depart from the choice \eqref{Nadef}, also without any loss of generality, and consider the choice $\Na=0$ 
in the context of \eqref{Sigmak1}.
The details of the wave propagation problem concerning the periodically repeating cell ${\mathscr{S}}_0$ are provided in Appendix \ref{wdisperse}\ref{appWmodeaway}; this is now important for the case of semi-infinite cracks since one possible mode of incidence corresponds to that from the cracked portions (with the assumption that the incidence from the infinite array still satisfies the Floquet condition \eqref{floquet}). 

After taking the Fourier transform along ${x}$, the general solution of the discrete Helmholtz equation \eqref{DH} for the scattered wave field in the lattice sites sandwiched between the two edges of ${\mathscr{S}}_0$, i.e., ${y}=0$ and ${y}={\mathtt{N}}-1$, is given by
${\su}^{{\mathrm{F}}}_{{y}}=a{\lambda}^{{y}}+b{\lambda}^{-{y}}, {y}\in{\ZZ}_0^{{\mathtt{N}}-1},$
where ${\lambda}$ is defined in \eqref{lambda}.
After solving for $a$ and $b$ in terms of ${\su}^{{\mathrm{F}}}_{0}, {\su}^{{\mathrm{F}}}_{{\mathtt{N}}-1}$, 
it is easy to see that
\begin{eqn}
{\su}^{{\mathrm{F}}}_{{y}}={\su}^{{\mathrm{F}}}_{0}\frac{{\lambda}^{{y}}-{\lambda}^{2{\mathtt{N}}-2-{y}}}{1-{\lambda}^{2{\mathtt{N}}-2}}+{\su}^{{\mathrm{F}}}_{{\mathtt{N}}-1}\frac{{\lambda}^{{\mathtt{N}}-1-{y}}-{\lambda}^{{\mathtt{N}}-1+{y}}}{1-{\lambda}^{2{\mathtt{N}}-2}}, {y}\in{\ZZ}_0^{{\mathtt{N}}-1}.
\label{uFy_k}
\end{eqn}
As observed in the previous section, the phase modulated periodicity \eqref{floquet} implies 
${\su}^{{\mathrm{F}}}_{{\mathtt{N}}-1}={\uppsi}{z}^{-{\mathtt{M}}}{\su}^{{\mathrm{F}}}_{-1}.$
We now consider the discrete Fourier transform as a sum of a pair of half-transforms, 
\begin{eqn}
u^{{\mathrm{F}}}(\varz)=u_+(\varz)+u_-(\varz), u_+(\varz)=\sum\nolimits_{m=0}^{+\infty}u_m\varz^{-m}, u_-(\varz)=\sum\nolimits_{m=-\infty}^{-1}u_m\varz^{-m}.
\label{defFT}
\end{eqn}
Employing the discrete Fourier transform \eqref{defFT} to the governing equation for a particle at 
$\yy=0$ and $\yy=-1$, respectively,
we get {(recall that $\HH$ is defined by \eqref{HHdef})}
\beqan
-{{b}^2}{{\mathit{v}}}^{\inc}_{0;-}({{z}})+{\su}_{0;-}({{z}})-{\su}_{-1;-}({{z}})=
({{\HH}}({{z}})+2){\su}^{{\mathrm{F}}}_{0}({{z}})-{\su}_{1}^{{\mathrm{F}}}({{z}})-{\su}_{-1}^{{\mathrm{F}}}({{z}}),\\
{{b}^2}{{\mathit{v}}}^{\inc}_{0;-}({{z}})+{\su}_{-1;-}({{z}})-{\su}_{0;-}({{z}})
=({{\HH}}({{z}})+2){\su}^{{\mathrm{F}}}_{-1}({{z}})-{\su}_{-2}^{{\mathrm{F}}}({{z}})-{\su}_{0}^{{\mathrm{F}}}({{z}}), 
{{z}}\in{{{\AAn}}},
\label{eqcrackgen0}\\
\text{where }
{{\mathit{v}}}^{\inc}_{0;-}({{z}})={b}^{-2}{{{\mathrm{A}}}}(1-{\exp}({i{\upkappa}_y}))\delta_{D-}({{z}} {{z}}_{{P}}^{-1}),
\label{v0nincF}
\text{and }
\delta_{D-}({{z}}){:=}\sum\nolimits_{n=-\infty}^{-1}{{z}}^{-n}, 
|{{z}}|<1,\\
{{z}}_{{P}}{:=} 
{\exp}({-i{\upkappa}\cos{\Theta}})\in{\mathbb{C}},
\text{further, \eqref{floquet} implies
${\uppsi}={z}_{{P}}^{{\mathtt{M}}}{\lambda}_{{P}}^{-{\mathtt{N}}}\text{ with }{\lambda}_{{P}}={\lambda}({z}_{{P}})
={\exp}({i{\upkappa}\sin{\Theta}}).$}
\label{phasefac2}
\eeqan
{In \eqref{v0nincF}${}_2$, it is clear that $\delta_{D-}({{z}})=z/(1-z)$ (using formula for the geometric series).}
Using \eqref{uFy_k} and 
\eqref{floquet}, the pair of coupled Wiener--Hopf equations \eqref{eqcrackgen0} can be expressed as 
\begin{eqn}
&\mathbf{{B}}\begin{bmatrix}{\su}_{0;+}\\{\su}_{-1;+}\end{bmatrix}+
(\mathbf{{B}}-\begin{bmatrix}1&-1\\-1&1\end{bmatrix})\begin{bmatrix}{\su}_{0;-}\\{\su}_{-1;-}\end{bmatrix}=\begin{bmatrix}-1\\1\end{bmatrix}{{b}^2}{{\mathit{v}}}^{\inc}_{0;-},\\
\hspace{-.1in}\text{where }
&\mathbf{{B}}=\begin{bmatrix}\nu_{{\mathtt{N}}}&\hspace{-.1in}-(1+{z}^{-{\mathtt{M}}}{\uppsi}\mu_{{\mathtt{N}}})\\-(1+{z}^{{\mathtt{M}}}{\uppsi}^{-1}\mu_{{\mathtt{N}}})&\nu_{{\mathtt{N}}}\end{bmatrix},
\nu_{{\mathtt{N}}}
=\frac{{\lambda}^{-{\mathtt{N}}}-{\lambda}^{{\mathtt{N}}}}{{\lambda}^{1-{\mathtt{N}}}-{\lambda}^{{\mathtt{N}}-1}}, 
\mu_{{\mathtt{N}}}
=\frac{{{\lambda}}^{-1}-{{\lambda}}}{{{\lambda}}^{1-{\mathtt{N}}}-{{\lambda}}^{{\mathtt{N}}-1}}. 
\label{numuK}
\end{eqn}
Although above problem \eqref{numuK} appears to be in the realm of matrix Wiener--Hopf kernels \cite{noble}, there exists a structure which leads to its reduction to scalar equation. 
Indeed, after addition of both component equations in \eqref{numuK}, it is found that
\begin{eqn}
{\su}_{-1;+}+{\su}_{-1;-}={\su}_{-1}^{{\mathrm{F}}}={\mathpzc{V}}{\su}_{0}^{{\mathrm{F}}}={\mathpzc{V}}({\su}_{0;+}+{\su}_{0;-}), 
\text{where }
{\mathpzc{V}}=-\frac{{\nu_{{\mathtt{N}}}-1}-\mu_{{\mathtt{N}}}{z}^{{\mathtt{M}}}{\uppsi}^{-1}}{{\nu_{{\mathtt{N}}}-1}-\mu_{{\mathtt{N}}}{z}^{-{\mathtt{M}}}{\uppsi}}.
\label{eqcrack4}
\end{eqn}
On the other hand, taking the difference of both equations \eqref{numuK} ({at this point} recall \eqref{scatterediff}),
using 
\begin{eqn}
\sv^{{\mathrm{F}}}=\sv_++\sv_-={\su}_{0}^{{\mathrm{F}}}-{\su}_{-1}^{{\mathrm{F}}}=(1-{\mathpzc{V}})\su_0^{{\mathrm{F}}}, 
\label{defvF}
\end{eqn}
and simplifying further {(following \cite{sWaveguide})}, 
a scalar Wiener--Hopf equation result in ${\sv}_{\pm}$, i.e.,
\begin{eqn}
{\sv}_{+}({{z}})+{{\sL}}_{{}}({{z}}) {\sv}_{-}({{z}})=-(1-{{\sL}}_{{}}({{z}}))\la^2{{\mathit{v}}}^{\inc}_{0;-}({{z}}), 
\quad{{z}}\in{{{\AAn}}}, 
\label{WHeqcrack}
\end{eqn}
\begin{eqn}
\text{where }
{{\sL}}_{{}}
&=\frac{{{\HH}}({z}){\mathtt{U}}_{{\mathtt{N}}-1}({{\vartheta}})}{2{\mathtt{T}}_{{\mathtt{N}}}({{\vartheta}})-({z}^{{\mathtt{M}}}{{\uppsi}}^{-1}+{z}^{-{\mathtt{M}}}{{\uppsi}})}=\frac{{\mathscr{N}}}{{\mathscr{D}}}.
\label{Lk_array}
\end{eqn}
In fact, by virtue of the presence of Chebyshev polynomials \cite{sCheby} of argument $\ctheta$ \eqref{defctheta}, an equivalent expression for kernel is ${{\sL}}_{{}}={\mathpzc{h}}^2\prod_{j=1}^{{\mathtt{N}}-1}({{\mathpzc{h}}}^2+4\sin^2\frac{1}{2}\frac{j\pi}{{\mathtt{N}}})/\big(\prod_{j=1}^{{\mathtt{N}}}({{\mathpzc{h}}}^2+4\sin^2\frac{1}{2}\frac{(j-\frac{1}{2})\pi}{{\mathtt{N}}})-({z}^{{\mathtt{M}}}{z}_{{P}}^{-{\mathtt{M}}}{\lambda}_{{P}}^{{\mathtt{N}}}+{z}^{-{\mathtt{M}}}{z}_{{P}}^{{\mathtt{M}}}{\lambda}_{{P}}^{-{\mathtt{N}}})\big)$;
recall that ${\uppsi}$ is defined by \eqref{floquet} and ${\lambda}_{{P}}, {z}_{{P}}$ by \eqref{phasefac2}.

The discrete Wiener--Hopf equation \eqref{WHeqcrack} has the same form as Eq. (2.23) 
in \cite{sK}
{and in fact it is almost identical to (2.8) in \cite{sWaveguide}}, hence, employing the same kind of, elementary, multiplicative factorization of the kernel ${{\sL}}_{{}}$ {(following the section \S2.4 of \cite{sWaveguide})}, {i.e., ${\sL}={\sL}_+{\sL}_-$ on $\AAn$}, we find
\begin{eqn}
{{\sL}}^{-1}_{{}+}({{z}}){\sv}_{+}({{z}})+{{{\sL}}_{{}}}_{-}({{z}}){\sv}_{-}({{z}})={{\mathpzc{C}}}({{z}}), 
\quad{{z}}\in{{{\AAn}}}_{{}}, \label{WH2}
\end{eqn}
where 
${{\mathpzc{C}}}({{z}})=({{\sL}}_{{}+}^{-1}({{z}})-{{{\sL}}_{{}}}_{-}({{z}})){{{\mathrm{A}}}}(1-{\exp}({i {\upkappa}_y}))\delta_{D-}({{z}} {{z}}_{{P}}^{-1}), {{z}}\in{{{\AAn}}}_{{}}.$
Further, an additive factorization of ${{\mathpzc{C}}}$, following \cite{sK}, is
\begin{eqn}
{{\mathpzc{C}}}={{\mathpzc{C}}}_{+}({{z}})+{{\mathpzc{C}}}_{-}({{z}}),
{{\mathpzc{C}}}_\pm({{z}})=\mp{{{\mathrm{A}}}}(1-{\exp}({i {\upkappa}_y}))({{\sL}}^{-1}_{{}+}({{z}}_{{P}})-{{{\sL}}_{{}}}_{\pm}^{\mp1}({{z}}))\delta_{D-}({{z}} {{z}}_{{P}}^{-1}), \quad{{z}}\in{{\AAn}}_{{}},
\label{Cpmk}
\end{eqn}
which
leads to
the exact solution of \eqref{WHeqcrack} as
\begin{eqn}
{\sv}_{\pm}({{z}})={{\mathpzc{C}}}_{\pm}({{z}}){{{{\sL}}_{{}}}}_{\pm}^{\pm1}({{z}}), {{z}}\in{\mathbb{C}}, |{{z}}|\gtrless\bfrac{\max}{\min}\{{{\mathit{R}}}_{\pm}, {{\mathit{R}}}_{L_{{}}}^{\pm1}\}.\label{vpm}\end{eqn}
Recall that ${z}_{{P}}$ is given by \eqref{phasefac2}. 

Using \eqref{defvF},
the exact solution \eqref{vpm} implies {($\sv^{{\mathrm{F}}}=\sv_++\sv_-$)}
\begin{eqn}
{\su}_{0}^{{\mathrm{F}}}=(1-{\mathpzc{V}})^{-1}\sv^{{\mathrm{F}}}, {\su}_{-1}^{{\mathrm{F}}}={\mathpzc{V}}(1-{\mathpzc{V}})^{-1}\sv^{{\mathrm{F}}} \text{ on }{{{\AAn}}}.
\label{uF0n1}
\end{eqn}
Along with \eqref{periodconditioncrackgen} and \eqref{uFy_k}, \eqref{uF0n1} provides the complete solution of the diffraction problem in integral form. 
By the inverse discrete Fourier transform (analogous to \eqref{fourierinvG}),
{the} displacement of the lattice 
at $({x}, 0)$ and {$(x, -1)$} is given by 
{
${\su}_{{x}, 0}=\frac{1}{2\pi i}\oint_{{{{{\mathcal{C}}}}}} \su_0^{\mathrm{F}}({{z}}){{z}}^{{x}-1}d{{z}}$ and ${\su}_{{x}, -1}=\frac{1}{2\pi i}\oint_{{{{{\mathcal{C}}}}}} \su_{-1}^{\mathrm{F}}({{z}}){{z}}^{{x}-1}d{{z}}$, respectively, with ${x}\in{\ZZ}$,}
where ${{\mathcal{C}}}$ is a rectifiable, closed, counterclockwise contour in the annulus ${{{\AAn}}}_{{}}$, and the remaining displacements at other edges are given by \eqref{periodconditioncrackgen}. 
{Analogous to ${\su}_{{x}, 0}$ and ${\su}_{{x}, -1}$,}
$\sv_{{x}}=\frac{1}{2\pi i}\oint_{{{{{\mathcal{C}}}}}} \sv^{{\mathrm{F}}}({{z}}){{z}}^{{x}-1}d{{z}}, {x}\in{\ZZ}, $
where $\sv^{{\mathrm{F}}}=\sv_++\sv_-$. {As} ${x}\to\pm\infty$, an 
{approximation} for $\sv$ can be obtained by analyzing \eqref{vpm} with
$\sv_{{x}}=\frac{1}{2\pi i}\oint_{{{{{\mathcal{C}}}}}} \sv_\pm({{z}}){{z}}^{{x}-1}d{{z}}, {x}\in{\ZZ}^{\pm}.$
Indeed, after deforming the contour of integration ${{{{{\mathcal{C}}}}}}$
{(expanding it to a circular contour with infinite radius when ${x}\to-\infty$ while contracting it to zero radius when ${x}\to+\infty$)} 
and applying residue calculus, {an approximation} is given by 
\begin{eqn}
\sv_{{x}}{\sim}\pm\sum\nolimits_{|{{z}}_{{\ast}}|\lessgtr1} \text{Res }{{\mathpzc{C}}}_{\pm}({{z}}_{{\ast}}){\mathpzc{L}}_{{}\pm}^{\pm1}({{z}}_{{\ast}}){{z}}_{{\ast}}^{{x}-1}, \quad\quad {x}\in{\ZZ}^{\pm}, {|{x}|\gg1},
\label{vm0asym1}
\end{eqn}
where the additive factors ${{\mathpzc{C}}}_{\pm}$ are given by \eqref{Cpmk}. 
{Let the unit circle in complex plane be denoted by $\T$.} 
{We} observe {that in} the limit ${\upomega}_2\to0+$ {(recall \eqref{damping})}, {while} considering $|{x}|\to\infty$, the approximation \eqref{vm0asym1} of $\sv_{{x}}$ {effectively} takes into account only the contributions due to {${{z}}_{{\ast}}$ approaching $\T$ appropriately and, evidently, representing} the outgoing wave modes, 
as expected \cite{brill}. 
{Also, observe that \eqref{Lk_array} allows the relations
${{\sL}}_+={\mathscr{N}}_+/{\mathscr{D}}_+$ and ${{\sL}}_-={\mathscr{N}}_-/{\mathscr{D}}_-$, using the factorization of numerator and denominator, i.e., ${\mathscr{N}}={\mathscr{N}}_+{\mathscr{N}}_-$ and ${\mathscr{D}}={\mathscr{D}}_+{\mathscr{D}}_-$, respectively, on $\AAn$.}
{Due to these observations, we} consider the following {natural} definitions, 
\begin{eqn}
{\mathcal{Z}^+}&=\{{z}\in\T \big| {{\mathscr{D}}_+({{z}})=0}\}, {\mathcal{Z}^-}=\{{z}\in\T \big| {{\mathscr{N}}_-({{z}})=0}\},
\label{Zer_sq_k}
\end{eqn}
{corresponding to outgoing waves towards the positive and negative ${x}$ axis away from the crack tip, respectively}.
To keep the same notation, the sets of ${z}$ belonging to
\begin{eqn}
\tilde{{\mathcal{Z}}}^{-}=\{{z}\in{\mathbb{T}} \big| {{\mathscr{D}}_-({{z}})=0}\}, \quad\quad
\tilde{{\mathcal{Z}}}^+=\{{z}\in{\mathbb{T}} \big| {{\mathscr{N}}_+({{z}})=0}\},
\label{Zer_sq_k_inc}
\end{eqn}
can be easily identified to those corresponding to the {{\em incident}} waves, namely, {{\em incident}} from the positive and negative ${x}$ axis {\em towards} the crack tip. Naturally, the case of incidence we are considering at the moment corresponds to those waves which travel from the bulk lattice so that ${z}_{{P}}\in\tilde{{\mathcal{Z}}}^{-}$. A little later we discuss the issue of incidence from the duct, i.e., corresponding 
{to}
$\tilde{{\mathcal{Z}}}^+$.\par
Coming back {to}
\eqref{vm0asym1}, after substitution of \eqref{Cpmk}, 
we find
\begin{eqn}
\sv_{{x}}&\sim{{{\mathrm{A}}}}{{a}}^{\inc}_{{}}\frac{{\mathscr{D}}_+({{z}}_{{P}})}{{\mathscr{N}}_+({{z}}_{{P}})}\sum\nolimits_{{{z}}_{{\ast}}\in{\mathcal{Z}^+}}\frac{{\mathscr{N}}_+({{z}}_{{\ast}})}{{\mathscr{D}}'_+({{z}}_{{\ast}})}\frac{{{z}}_{{\ast}}^{{x}}}{{{z}}_{{\ast}}-{{z}}_{{P}}}, {x}\to+\infty,\\
\sv_{{x}}&\sim{{{\mathrm{A}}}}{{a}}^{\inc}_{{}}(-{{z}}_{{P}}^{{x}}+\frac{{\mathscr{D}}_+({{z}}_{{P}})}{{\mathscr{N}}_+({{z}}_{{P}})}\sum\nolimits_{{{z}}_{{\ast}}\in{\mathcal{Z}^-}}\frac{{\mathscr{D}}_-({{z}}_{{\ast}})}{{\mathscr{N}}'_-({{z}}_{{\ast}})}\frac{{{z}}_{{\ast}}^{{x}}}{{{z}}_{{\ast}}-{{z}}_{{P}}}), {x}\to-\infty,
\label{vm0asym}
\end{eqn}
\begin{eqn}
\text{where 
${{a}}^{\inc}_{{}}$ is given by }
{{a}}^{\inc}_{{}}{:=}1-{\exp}({i{\upkappa}_y}).
\label{avinc_k}
\end{eqn}

\begin{remark}
{In}
\eqref{vm0asym} corresponding to ${x}\to+\infty$, ${{z}}_{{P}}^{-1}$ is included in sum (here ${\mathscr{D}}_+({{z}}_{{P}}^{-1})=0, {\mathscr{D}}_-({{z}}_{{P}}^{-1})\ne0$ but ${\mathscr{D}}_-({{z}}_{{P}})=0$). 
{In}
\eqref{vm0asym} for ${x}\to-\infty$, ${{z}}_{{P}}$ does not occur in sum (${\mathscr{D}}_-({{z}}_{{P}})=0, {\mathscr{D}}_+({{z}}_{{P}})\ne0$ but ${\mathscr{D}}_+({{z}}_{{P}}^{-1})=0$). 
A testimony to preceding sentences. 
\label{remzpn}\end{remark}
Using \eqref{vm0asym}, after simplification, the total field is given by $\sv^{{t}}_{{x}}=\sv^{\inc}_{{x}}+\sv_{{x}}$. 
Thus, for a wave incident from the bulk lattice, i.e. region (1) of Fig. \ref{squarelatticearraycracks} in front of the staggered array, the transmitted waves includes the contribution from the residue associated with the incident wave which cancels another equal contribution, while the reflected waves includes the contribution from the residue associated with the reflected wave determined by the incident wavenumber. 

Using \eqref{vm0asym}, we can also construct the asymptotic expansion of scattered wave field. Note that the far-field can be determined 
in terms of the (propagating) normal modes associated with the two different portions (ahead, indicated by subscript $+$, and behind, by $-$).
The normal modes for a square lattice waveguide with fixed or free boundary are known (see also \cite{sCheby}).
Let 
\begin{eqn}
{{z}}_{{\ast}}={\exp}({-i{\upxi}}), {\lambda}({z}_{{\ast}})={\exp}({i{\upeta}}), \text{ and }{{\kappa}}_{{z}_{{\ast}}}\text{ refer to specific normal mode depending on }{{z}}_{{\ast}}.
\label{zxidefn}
\end{eqn} 
{Thus, (with ${J}_{{\kappa}}$ as amplitudes of relevant wave modes)}
\begin{eqn}
\su_{{x}, {y}}&\sim{{{\mathrm{A}}}}\sum\nolimits_{{{z}}_{{\ast}}\in{\mathcal{Z}^+}}{J}_{{{\kappa}}_{{z}_{{\ast}}}}{{a}}_{+({{\kappa}}_{{z}_{{\ast}}}){{y}}}{{z}}_{{\ast}}^{{x}}, {x}\to+\infty, \\
\su_{{x}, {y}}&\sim{{{\mathrm{A}}}}\sum\nolimits_{{{z}}_{{\ast}}\in{\mathcal{Z}^-}}{J}_{{{\kappa}}_{{z}_{{\ast}}}}{{a}}_{-({{\kappa}}_{{z}_{{\ast}}}){{y}}}{{z}}_{{\ast}}^{{x}}-\mathrm{A}{J}_{{{\kappa}}^{\inc}}{{a}}_{({{{\kappa}}^{\inc}}){{y}}}{{z}}_{{P}}^{{x}}, {x}\to-\infty,
\label{assumedfarfield_k_sq}
\end{eqn}
respectively, so that the total field is the desired one. 
The expression \eqref{assumedfarfield_k_sq} also yields $\sv_{{x}}$ as ${x}\to\pm\infty$.
Note that the wave modes ahead of the array are given by \eqref{eigvaleqn}, \eqref{eigveceqn}, and \eqref{eigvecnorm}, while those behind the scattering edges are given by \eqref{oneDfree}.
Evidently, this leads to 
the total displacement field 
\begin{eqn}
\su^{{t}}_{{x}, {y}}&\sim
{\su}_{{x}, {y}}^{\inc}
+{{{\mathrm{A}}}}\frac{{\mathscr{D}}_+({{z}}_{{P}})}{{\mathscr{N}}_+({{z}}_{{P}})}\sum\nolimits_{{{z}}_{{\ast}}\in{\mathcal{Z}^+}}\frac{{{a}}^{\inc}_{{}}{{a}}_{+({{\kappa}}_{{z}_{{\ast}}}){{y}}}{{z}}_{{\ast}}^{{x}}}{{{a}}_{+({{\kappa}}_{{z}_{{\ast}}})0}-{\uppsi}^{-1}{z}_{{\ast}}^{{\mathtt{M}}}{{a}}_{+({{\kappa}}_{{z}_{{\ast}}}){{\mathtt{N}}}-1}}\frac{1}{{{z}}_{{\ast}}-{{z}}_{{P}}} \frac{{\mathscr{N}}_+({{z}}_{{\ast}})}{{\mathscr{D}}'_+({{z}}_{{\ast}})}\\
\su^{{t}}_{{x}, {y}}&\sim{{{\mathrm{A}}}}\frac{{\mathscr{D}}_+({{z}}_{{P}})}{{\mathscr{N}}_+({{z}}_{{P}})}\sum\nolimits_{{{z}}_{{\ast}}\in{\mathcal{Z}^-}}\frac{{{a}}^{\inc}_{{}}{{a}}_{-({{\kappa}}_{{z}_{{\ast}}}){{y}}}{{z}}_{{\ast}}^{{x}}}{{{a}}_{-({{\kappa}}_{{z}_{{\ast}}})0}-{\uppsi}^{-1}{z}_{{\ast}}^{{\mathtt{M}}}{{a}}_{-({{\kappa}}_{{z}_{{\ast}}}){\mathtt{N}}-1}}\frac{1}{{{z}}_{{\ast}}-{{z}}_{{P}}} \frac{{\mathscr{D}}_-({{z}}_{{\ast}})}{{\mathscr{N}}'_-({{z}}_{{\ast}})},
\label{farfield_k_sq}
\end{eqn}
as ${x}\to+\infty$ and ${x}\to-\infty$, respectively. 
In \eqref{vm0asym} and \eqref{farfield_k_sq}, the substitution of ${\mathpzc{L}_{{\mathtt{N}}}}_{{}}$ by 
${\mathscr{N}}/{\mathscr{D}}$ has been used; in this context, 
${{\mathscr{N}}_+}/{{\mathscr{D}}'_+}={{\mathscr{D}}_-{\mathscr{N}}_+}/{{\mathscr{D}}'}, {{\mathscr{D}}_-}/{{\mathscr{N}}'_-}={{\mathscr{D}}_-{\mathscr{N}}_+}/{{\mathscr{N}}'}, \text{ etc}.$
With details relegated to Appendix \ref{RTsemi},
the transmittance, i.e., the energy flux transmitted into the cracked portion per unit incident energy flux, is given by
\begin{eqn}
{{\mathcal{T}}}&=\frac{{{z}}_{{P}}{\mathscr{N}}_-({{z}}_{{P}}){\mathscr{D}}_+({{z}}_{{P}})}{\overline{{\mathscr{D}}'_-({{z}}_{{P}})}\overline{{\mathscr{N}}_+({{z}}_{{P}})}}\sum\nolimits_{{{z}}\in{\mathcal{Z}^-}}\frac{\overline{{\mathscr{D}}_-({{z}}){\mathscr{N}}_+({{z}})}}{{\mathscr{N}}'_-({{z}}){\mathscr{D}}_+({{z}})}\frac{{{z}}_{{P}}}{({{z}}-{{z}}_{{P}})^2}.
\label{Trans_k_sq_Psemi}
\end{eqn}
Analogous result holds for the reflectance ${{\mathcal{R}}}$ \eqref{Ref_k_sq_P}.
This is a special form of expression for ${{\mathcal{R}}}$ and ${{\mathcal{T}}}$ as it coincides with 
{that}
for the bifurcated waveguides \cite{sWaveguide}.

\begin{figure}[h]
\centering
\includegraphics[width=.4\textwidth]{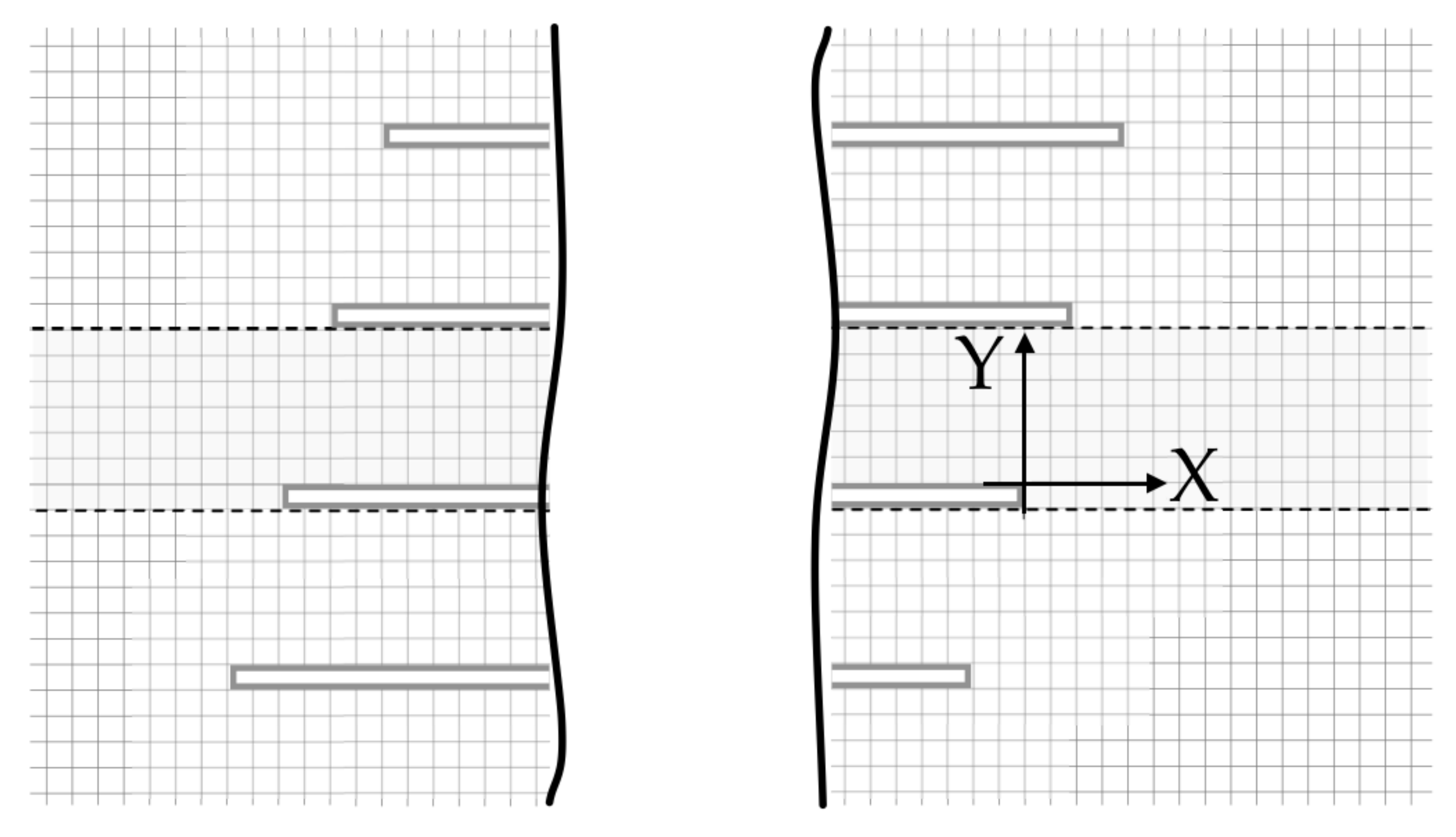}
\caption{{An} illustration of an infinite array of ``long'' but finite cracks with interaction between the two arrays of crack tips.}
\label{energybalanceacrosstwoedgesK}
\end{figure}

The {analysis} of ${{\mathcal{R}}}$ and ${\mathcal{T}}$ for finite cracks {is} provided in Appendix \ref{RT}. Here, ${{\mathcal{R}}}$ (resp. ${\mathcal{T}}$) refers to the energy flux reverted back to the same side as that of the incident wave while treating the infinite array of finite-length cracks as a kind of {\em interface}. However, when $\Nb\gg1$ it is natural to resolve the issue of energy flux transmission for the waves generated inside the cracks and for this purpose Fig. \ref{energybalanceacrosstwoedgesK} gives a helpful hint. The following {deals} with this aspect of semi-infinite cracks.

\subsection*{Wave incidence from the ducts}
In order to maintain the convenience of Floquet periodicity,
it is assumed apriori that steady state has been reached for an incident wave field that arrives from all of the infinite number of ducts and that the scattered waves are excited by all the edges. This assumption is derived from the scenario presented in Fig. \ref{energybalanceacrosstwoedgesK}; employed to tackle the issue of finite cracks when $\Nb\gg1$ by posing the Wiener--Hopf problem for two kinds of incidence.
The scattering in this case of {\em wave incidence from the ducts} occurs due to the intact bonds ahead of the staggered array of defects. A schematic illustration is provided in Fig. \ref{squarelatticearraycracks}.
Analogous to \eqref{uinc}, it is assumed that the incident wave {is}
\begin{eqn}
{\su}_{{x}, {y}}^{\inc}{:=}{{{\mathrm{A}}}}{{a}}_{({{\kappa}}^{\inc}){{\nu}}}{P}^{\lfloor{y}/{\mathtt{N}}\rfloor}{\exp}({i{\upkappa}_x {x}-i{\omega} t}), \quad({x}, {y})\in{{\ZZ\times\ZZ_{{\mathtt{N}}}}}, {\upkappa}_x>0, {\nu}=\text{mod}({y}, {\mathtt{N}}),
\label{uinc_incsca}
\end{eqn}
where ${{a}}_{({{\kappa}}^{\inc})}$ denotes {wave} mode in any of the portions {\em between} the scattering edges and ${P}$ is 
the phase factor. 
Note that the phase factor in the duct located between ${y}=j{\mathtt{N}}$ and ${y}=j{\mathtt{N}}+{\mathtt{N}}-1$ is ${P}^j$. 
Since $|{P}|=1$, let ${P}={\exp}({i{\upkappa}^{{\#}}_y{\mathtt{N}}})$ for some ${\upkappa}^{{\#}}_y\in[-\pi, \pi].$ Thus, the incident wave imposes a Floquet--Bloch multiplier ${\exp}({i{\upkappa}_x{\mathtt{M}}}+{i{\upkappa}^{{\#}}_y{\mathtt{N}}})$; recall \eqref{floquet}.

\begin{remark}
 Indeed, ${\su}^{\inc}_{{x}+{\mathtt{M}}, {y}+{\mathtt{N}}}={\uppsi}{\su}^{\inc}_{{x}, {y}}, {y}\in{\ZZ}_0^{{\mathtt{N}}-1}, {x}\in{\ZZ},$
where 
\begin{eqn}
{\uppsi}={\exp}({i{\upkappa}_x{\mathtt{M}}+i{\upkappa}^{{\#}}_y}{\mathtt{N}}).
\label{phasefac_incsca}
\end{eqn}
Given the wave mode ${{\kappa}}^{\inc}$ in the duct portion, the frequency ${\upomega}$ and incident wave number ${\upkappa}_x$ are related by the duct dispersion relation. Thus, the scattering of a specific duct mode involves two free parameters ${\upkappa}_x$ and ${\upkappa}^{{\#}}_y$ where the former yields a specific frequency in ${{\kappa}}^{\inc}$ mode (similar to the case of incidence from the bulk lattice where ${\upkappa}_y$ and ${\upkappa}_x$ can be chosen arbitrarily
and ${\upomega}$ is provided by the square lattice dispersion relation). 
{It is useful to note that ${\upkappa}_y$ is pre-determined by the bulk incidence while ${\upkappa}_x$ depends on the transmitted wave numbers towards the `other' edge.}
\end{remark}

Taking into account the intact bonds between ${y}=0$ and ${y}=-1$ for ${x}\ge0$, and also ${y}=0+{\mathtt{N}}$ and ${y}=-1+{\mathtt{N}}$ for ${x}\ge{\mathtt{M}}$ and so on, and the broken nature of all other bonds, the equation that must be satisfied by ${\su}$ at ${y}=0$ is found to be, in terms of \eqref{numuK}, 
\begin{eqn}
{\la^2}{{\mathit{v}}}^{\inc}_{0;+}&=\nu_{{\mathtt{N}}}{\su}_{0;+}-(1+{z}^{-{\mathtt{M}}}{\uppsi}\mu_{{\mathtt{N}}}){\su}_{-1;+}+(\nu_{{\mathtt{N}}}-1){\su}_{0;-}-{z}^{-{\mathtt{M}}}{\uppsi}\mu_{{\mathtt{N}}}{\su}_{-1;-},\\
-{\la^2}{{\mathit{v}}}^{\inc}_{0;+}&=\nu_{{\mathtt{N}}}{\su}_{-1;+}-(1+{z}^{{\mathtt{M}}}{\uppsi}^{-1}\mu_{{\mathtt{N}}}){\su}_{0;+}+(\nu_{{\mathtt{N}}}-1){\su}_{-1;-}-{z}^{{\mathtt{M}}}{\uppsi}^{-1}\mu_{{\mathtt{N}}}{\su}_{0;-},
\label{eqcrack2_2_incsca}
\end{eqn}
\begin{eqn}
\text{where }
{{\mathit{v}}}^{\inc}_{0;+}
&=\la^{-2}{{{\mathrm{A}}}}{{a}}^{\inc}_{{}}\delta_{D+}({{z}} {{z}}_{{P}}^{-1}), {{z}}_{{P}}{:=} {\exp}({i{\upkappa}_x})\in{\mathbb{C}},
\label{v0nincF_incsca}
{{a}}^{\inc}_{{}}{:=}{{a}}_{({{{\kappa}}^{\inc}})0}-{\uppsi}^{-1}{z}^{{\mathtt{M}}}_{{P}}{{a}}_{({{{\kappa}}^{\inc}}){\mathtt{N}}-1}.
\end{eqn}
Indeed, after addition of both equations, it is found that \eqref{eqcrack4} holds, and by taking the difference of both equations,
using \eqref{defvF},
and simplifying further, the resulting 
equation is found to be
\begin{eqn}
{\sv}_{+}({{z}})+{{\sL}}_{{}}({{z}}) {\sv}_{-}({{z}})=(1-{{\sL}}_{{}}({{z}}))\la^2{{\mathit{v}}}^{\inc}_{0;+}({{z}}), 
\quad\quad{{z}}\in{{{\AAn}}}, 
\label{WHeqcrack_incsca}
\end{eqn}
which is the scalar discrete Wiener--Hopf equation for $\sv$, as desired, where ${{\sL}}_{{}}$ is given by \eqref{Lk_array}.
Note that {as} ${{z}}_{{P}}={\exp}({i{\upkappa}_x})$, for the dissipative case {it is found that} $|{{z}}_{{P}}|<1.$ 
Above can be compared and contrasted with the case of incident wave from the bulk lattice, i.e. the intact part of the waveguide, \eqref{WHeqcrack} and some relations between involved entities can be found; for instance, the Wiener--Hopf kernel remains same. 
\eqref{WHeqcrack_incsca} {has} same form as \eqref{WHeqcrack}, hence \eqref{WH2} {holds} with
\begin{eqn}
{{\mathpzc{C}}}({{z}})=-({{\sL}}_{{}+}^{-1}({{z}})-{{{\sL}}_{{}}}_{-}({{z}})){{{\mathrm{A}}}}{{a}}^{\inc}_{{}}\delta_{D+}({{z}} {{z}}_{{P}}^{-1}), {{z}}\in{{{\AAn}}}_{{}}, 
\label{Cz_incsca}
\delta_{D+}({{z}}){:=}\sum\nolimits_{n=0}^{+\infty}{{z}}^{-n}, 
|{{z}}|>1.
\end{eqn}
An additive factorization, ${{\mathpzc{C}}}={{\mathpzc{C}}}_{+}({{z}})+{{\mathpzc{C}}}_{-}({{z}}),$ is constructed 
with
\begin{eqn}
{{\mathpzc{C}}}_\pm({{z}})&=\pm{{{\mathrm{A}}}}{{a}}^{\inc}_{{}}({{{\sL}}_{{}}}_{-}({{z}}_{{P}})-{{{\sL}}_{{}}}_{\pm}^{\mp1}({{z}}))\delta_{D+}({{z}} {{z}}_{{P}}^{-1}), \quad\quad{{z}}\in{{\AAn}}_{{}}.
\label{Cpmk_incsca}
\end{eqn}
Finally, in terms of the one-sided discrete Fourier transform, 
$\sv^{{\mathrm{F}}}$ is given by \eqref{vpm}, {while ${\su}_{0}^{{\mathrm{F}}}$ and ${\su}_{-1}^{{\mathrm{F}}}$ are} given by \eqref{uF0n1}.

Coming over now to the question of asymptotic approximation of the solution deep into the portions away from the crack tips.
Using \eqref{Cpmk_incsca}, 
\begin{eqn}
\sv_{{x}}&\sim{{{\mathrm{A}}}}{{a}}^{\inc}_{{}}(-{{z}}_{{P}}^{{x}}+\frac{{\mathscr{N}}_-({{z}}_{{P}})}{{\mathscr{D}}_-({{z}}_{{P}})}\sum\nolimits_{{{z}}_{{\ast}}\in{\mathcal{Z}^+}}\frac{{\mathscr{N}}_+({{z}}_{{\ast}})}{{\mathscr{D}}'_+({{z}}_{{\ast}})}\frac{{{z}}_{{\ast}}^{{x}}}{{{z}}_{{\ast}}-{{z}}_{{P}}}), {x}\to+\infty,\\
\sv_{{x}}&\sim{{{\mathrm{A}}}}{{a}}^{\inc}_{{}}\frac{{\mathscr{N}}_-({{z}}_{{P}})}{{\mathscr{D}}_-({{z}}_{{P}})}\sum\nolimits_{{{z}}_{{\ast}}\in{\mathcal{Z}^-}}\frac{{\mathscr{D}}_-({{z}}_{{\ast}})}{{\mathscr{N}}'_-({{z}}_{{\ast}})}\frac{{{z}}_{{\ast}}^{{x}}}{{{z}}_{{\ast}}-{{z}}_{{P}}}, {x}\to-\infty.
\label{vm0asym_incsca}
\end{eqn}
It is observed in the {above case}
corresponding to ${x}\to+\infty$
{that}
${{z}}_{{P}}$ does not occur in the sum (here ${\mathscr{N}}_+({{z}}_{{P}})=0, {\mathscr{N}}_-({{z}}_{{P}})\ne0$ but ${\mathscr{N}}_-({{z}}_{{P}}^{-1})=0$) as anticipated. 
Recall \eqref{Zer_sq_k} and \eqref{Zer_sq_k_inc} 
{for the definitions of ${\mathcal{Z}}^\pm$}.
In the expression for ${x}\to-\infty$, ${{z}}_{{P}}^{-1}$ is included in the sum (${\mathscr{N}}_+({{z}}_{{P}}^{-1})\ne0, {\mathscr{N}}_-({{z}}_{{P}}^{-1})=0$ but ${\mathscr{N}}_+({{z}}_{{P}})=0$). 

Analogous to \eqref{farfield_k_sq}, resulting from \eqref{vm0asym_incsca}, the total displacement field is written as
\begin{eqn}
\su^{{t}}_{{x}, {y}}&\sim{{{\mathrm{A}}}}\frac{{\mathscr{N}}_-({{z}}_{{P}})}{{\mathscr{D}}_-({{z}}_{{P}})}\sum\nolimits_{{{z}}\in{\mathcal{Z}^+}}\frac{{{a}}^{\inc}_{{}}{{a}}_{+({{\kappa}}_{{z}}){{y}}}{{z}}^{{x}}}{{{a}}_{+({{\kappa}}_{{z}})0}-{\uppsi}^{-1}{z}^{{\mathtt{M}}}{{a}}_{+({{\kappa}}_{{z}}){{\mathtt{N}}}-1}}\frac{1}{{{z}}-{{z}}_{{P}}} \frac{{\mathscr{N}}_+({{z}})}{{\mathscr{D}}'_+({{z}})}\\
\su^{{t}}_{{x}, {y}}&\sim{{{\mathrm{A}}}}{{a}}_{({{{\kappa}}^{\inc}}){{y}}}{{z}}_{{P}}^{{x}}+{{{\mathrm{A}}}}\frac{{\mathscr{N}}_-({{z}}_{{P}})}{{\mathscr{D}}_-({{z}}_{{P}})}\sum\nolimits_{{{z}}\in{\mathcal{Z}^-}}\frac{{{a}}^{\inc}_{{}}{{a}}_{-({{\kappa}}_{{z}}){{y}}}{{z}}^{{x}}}{{{a}}_{-({{\kappa}}_{{z}})0}-{\uppsi}^{-1}{z}^{{\mathtt{M}}}{{a}}_{-({{\kappa}}_{{z}}){\mathtt{N}}-1}}\frac{1}{{{z}}-{{z}}_{{P}}} \frac{{\mathscr{D}}_-({{z}})}{{\mathscr{N}}'_-({{z}})},
\label{farfield_k_sq_incsca}
\end{eqn}
as ${x}\to+\infty$ and ${x}\to-\infty$, respectively. 
Finally, the transmittance, i.e., the energy flux transmitted into the intact portion per unit incident energy flux from the cracked portion, is given by
\begin{eqn}
{{\mathcal{T}}}&=\frac{({{v}}({\upxi}_{{P}}))^{-1}}{|{{\mathpzc{L}_{{\mathtt{N}}}}}_{{}-}^{-1}({{z}}_{{P}})|^2}\sum\nolimits_{{{z}}\in{\mathcal{Z}^+}}\frac{|{{a}}^{\inc}_{{}}|^2}{-\frac{{\mathscr{N}}({z})}{{\mathtt{N}}{\mathtt{U}}_{{\mathtt{N}}-1}({\vartheta})}}\frac{1}{\overline{{z}}-\overline{{z}}_{{P}}} \frac{\overline{{\mathscr{D}}_-({{z}})}\overline{{\mathscr{N}}_+({{z}})}}{\overline{{\mathscr{D}}'({{z}})}}\frac{1}{{{z}}-{{z}}_{{P}}} \frac{{\mathscr{N}}({{z}})}{{\mathscr{D}}'_+({{z}}){\mathscr{N}}_-({{z}})}{{v}}({\upxi})\\&=\frac{1}{2}{i}\frac{{\upomega}^{-1}|{{a}}^{\inc}_{{}}|^2}{({{v}}({\upxi}_{{P}}))|{{\mathpzc{L}_{{\mathtt{N}}}}}_{{}-}^{-1}({{z}}_{{P}})|^2}\sum\nolimits_{{{z}}\in{\mathcal{Z}^+}}\frac{\overline{{\mathscr{D}}_-({{z}}){\mathscr{N}}_+({{z}})}}{{\mathscr{D}}'_+({{z}}){\mathscr{N}}_-({{z}})}\frac{{{z}}_{{P}}}{({{z}}-{{z}}_{{P}})^2},
\label{Trans_k_sq_P_incsca}
\end{eqn}
while, the reflectance is given by
\begin{eqn}
\hspace{-.1in}{{\mathcal{R}}}&=\frac{({{v}}({\upxi}_{{P}}))^{-1}}{|{{\mathpzc{L}_{{\mathtt{N}}}}}_{{}-}^{-1}({{z}}_{{P}})|^2}\sum\nolimits_{{{z}}\in{\mathcal{Z}^-}}\frac{{{2{\mathtt{N}}}}|{{a}}^{\inc}_{{}}|^2}{\cos{\mathtt{N}}{{\upeta}_{{\kappa}}}{({\HH}({z})+4)}{\mathscr{D}}({{z}})}\frac{{\mathscr{D}}({{z}}){|{{v}}({\upxi})|}}{\overline{{\mathscr{N}}'({{z}})}}\frac{\overline{{\mathscr{D}}_-({{z}}){\mathscr{N}}_+({{z}})}}{{\mathscr{N}}'_-({{z}}){\mathscr{D}}_+({{z}})}\frac{1}{\overline{{z}}-\overline{{z}}_{{P}}}\frac{1}{{{z}}-{{z}}_{{P}}} \\&=\frac{1}{2}{i}\frac{{\upomega}^{-1}|{{a}}^{\inc}_{{}}|^2}{({{v}}({\upxi}_{{P}}))|{{\mathpzc{L}_{{\mathtt{N}}}}}_{{}-}^{-1}({{z}}_{{P}})|^2}\sum\nolimits_{{{z}}\in{\mathcal{Z}^-}}\frac{\overline{{\mathscr{D}}_-({{z}}){\mathscr{N}}_+({{z}})}}{{\mathscr{N}}'_-({{z}}){\mathscr{D}}_+({{z}})}\frac{{{z}}_{{P}}}{({{z}}-{{z}}_{{P}})^2},
\label{Ref_k_sq_P_incsca}
\end{eqn}
(recall \eqref{Trans_k_sq_P} and \eqref{Ref_k_sq_P}, respectively).
The coefficient in front of the sum can be simplified as 
\begin{eqn}
\frac{1}{2}{\upomega}^{-1}{i}\frac{|{{a}}^{\inc}_{{}}|^2}{({{v}}({\upxi}_{{P}}))|{{\mathpzc{L}_{{\mathtt{N}}}}}_{{}-}^{-1}({{z}}_{{P}})|^2}
&=\frac{{{z}}_{{P}}{\mathscr{N}}_-({{z}}_{{P}}){\mathscr{D}}_+({{z}}_{{P}})}{\overline{{\mathscr{D}}_-({{z}}_{{P}})}\overline{{\mathscr{N}}'_+({{z}}_{{P}})}}.
\label{CRexp_incsca}
\end{eqn}
The detailed derivation is omitted in the main article.

\section{Numerical Results}
\label{numresult}

A numerical scheme on the lines of that stated in the Appendix of \cite{sWaveguide}, for bifurcated waveguides, has been used to solve directly the discrete scattering problem involving the array of semi-infinite cracks as well as finite cracks on the lattice. 
We omit the graphical results in the main paper but remark that the corresponding results have been found to be in excellent agreement 
with the semi-analytical solution of \S\ref{reduce}.
It is also found that when the separation between the adjacent cracks, $\NN,$ is large, the solution near the edge of any crack in the array allows an approximation by that for a single crack on a square lattice, modulo a suitable phase factor. 
\begin{figure}[ht]
\centering
\includegraphics[width=\textwidth]{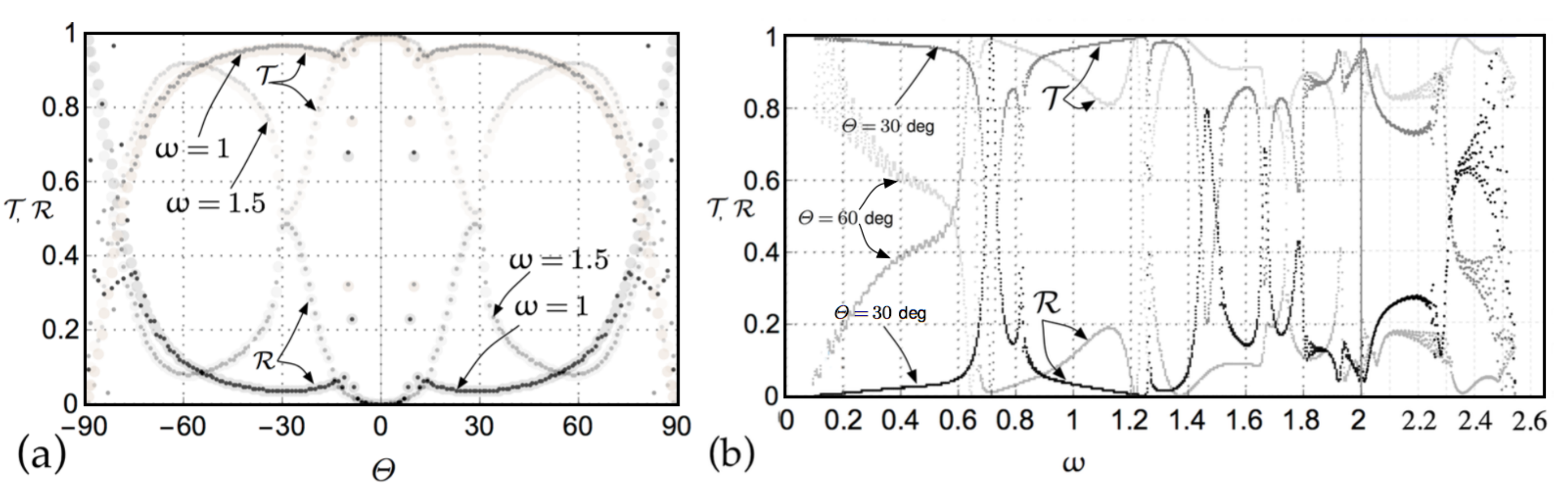}
\caption{Reflectance ${{\mathcal{R}}}$ and transmittance ${\mathcal{T}}$ have been plotted against the incident angle $\Theta$ {in (a) 
and against the frequency $\oo$ in (b)}. 
The semi-analytical results are plotted in {lighter shade and bigger dots} whereas the numerical results are plotted in black and grey colours. 
{The parameters used for these figures are $\MM=0, \NN=5, \Nb=5$.}}
\label{energyb}
\end{figure}
{The main results of this paper concern the aspects of transmission of energy.}
{The reflected (resp. transmitted) energy flux per unit incident energy flux, called reflectance (resp. transmittance) are calculated numerically using an analytical expression has been derived in Appendix \ref{RT}.}

{In Fig. \ref{energyb}(a), the} reflectance (${{\mathcal{R}}}$, see \eqref{reflectance}${}_1$) and transmittance (${\mathcal{T}}$, see \eqref{reflectance}${}_2$) have been plotted against the incident angle $\Theta$. It can be verified that the sum of the reflectance and transmittance is one, which is the consequence of the balance of the mechanical energy. 
It can also be seen that most of the energy is transmitted while a very small amount is reflected for certain choices of $\Theta$. Such information is anticipated to be useful for the planning and engineering of nanostructures where the high frequency scattering plays a major role. 
Fig. \ref{energyb}(a) gives the comparison of the semi-analytical and numerical results also; the two approaches show a good agreement. 

{In Fig.} \ref{energyb}(b), the reflectance and transmittance versus the frequency $\oo$ {is shown}. When the incident wave frequency 
is near zero, the discrete solution approaches that of its continuous counterpart \cite{sK}. The physical effects of discreteness become visible for much higher frequencies belonging to the passband, 
see the portions lying on the right side of the plots shown in Fig. \ref{energyb}(b).
As part of the analysis of some key features, note that in Fig. \ref{energyb}(b) there are numerous peaks or valleys in the transmittance and reflectance. 
In Fig. \ref{energyb}(a) and \ref{energyb}(b), 
the numerical oscillations 
depend upon the domain of numerical calculations. The large domain fixes the oscillations resulting into smooth curves. Such oscillations are absent in the semi-analytical results as can be seen in Fig. \ref{energyb}(a). 
{
The limit of $\oo$ value in Fig. \ref{energyb}(b) corresponds to those frequencies which lie in the fundamental zone, i.e., $(\kk\cos\Theta, \kk\sin\Theta)\in[-\pi, \pi]^2$ in \eqref{uinc}.}

{The} transmittance (calculated numerically, while an analytical expression has been derived in Appendix \ref{RT}) is illustrated further in Fig. \ref{abs}. The transmittance 
versus the frequency (in the passband with {the limit of $\oo$ value in Fig. \ref{abs}(a)--(d) corresponding to that which lies in the fundamental zone such that $(\kk\cos\Theta, \kk\sin\Theta)\in[-\pi, \pi]^2$ in \eqref{uinc}}) of the incident wave for {some values of the relevant} parameters, namely, incident angle {relative to normal to the crack tips} (denoted by ${\upbeta}$ \eqref{normalincang}), spacing between two consecutive cracks $\NN$, stagger between cracks $\MM$, and crack length $\Nb$ has been shown in Fig. \ref{abs}. When $\MM=0$, that is {shown as black curve in Fig. \ref{abs}(a)}, when the cracks are not offset with each other, the incident wave at normal incidence (${\upbeta}=0^\circ$) is transmitted perfectly without getting scattered for all frequencies in the passband. The complete transmission was also observed in the case of plates with periodically arranged parallel rectangular slots carrying a longitudinal elastic wave at normal incidence as reported in
\cite{su2016} (see Figs 3 and 4). The low frequency region of Fig. \ref{abs}{(a)} represents the solution (for $\MM=0$) which matches with that of the continuous counterpart presented 
in \cite{su2016}. It is not surprising to see that the behaviour is extended for much higher frequencies since it is expected from the assumed simplified model. 
\begin{figure}[ht!]
\vspace{-0mm}
\centering
\includegraphics[width=\textwidth]{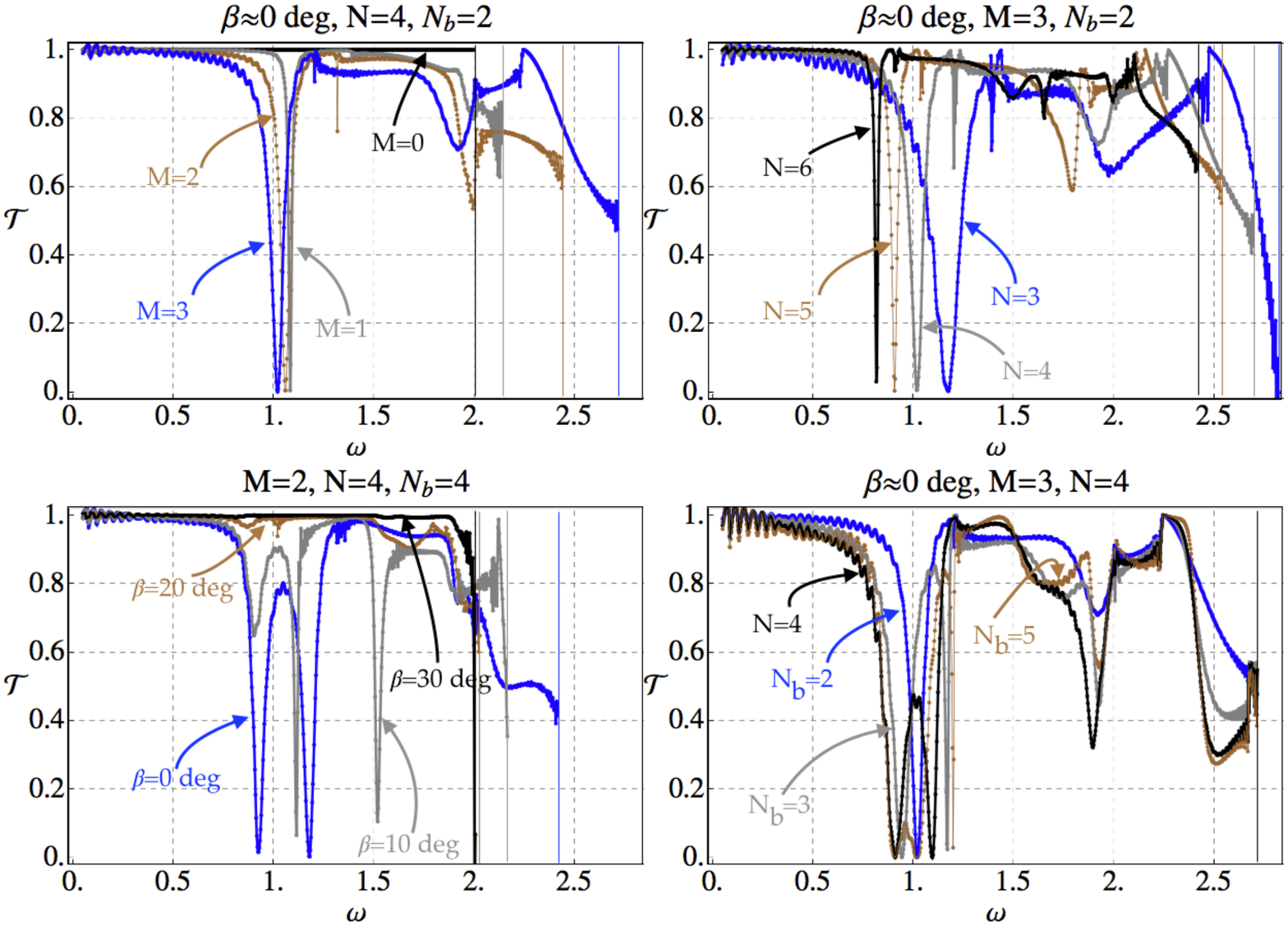}
\caption{{Transmittance ${\mathcal{T}}$ (vertical axis) versus frequency $\omega$ (horizontal axis) 
for parameters shown.}}
\label{abs}
\end{figure}
High transmission can be also seen, at normal incidence, when the cracks are staggered (i.e., when $\MM\neq0$). However, in {the presence of stagger}, there are certain frequencies in the passband at which the incident waves are {(almost)} completely reflected (as indicated by dip(s) in the transmission curves {in all cases of Fig. \ref{abs}(a)--(d)}). 
{Note that the oscillations present near $\oo=0$ in Fig. \ref{abs}(a)--(d) are numerical figments and disappear with increase in numerical domain size simulating the infinite lattice}.
{It can be observed that } 
the range of the frequencies reflected back decreases, i.e., the band of frequency reflected shrinks in its width{,} as 
{the inter-crack spacing} $\NN$ increases, {see Fig. \ref{abs}{(b)}}. 
{At oblique incidence (${\upbeta}\neq0^\circ$) there are no frequencies in the passband at which the complete reflection or transmission occurs as illustrated by the black curve for ${\upbeta}=30^\circ$ in Fig. \ref{abs}(c).}
The number of dips{, i.e., frequencies corresponding to a completely reflected character, increases as the crack length $\Nb$ is increased as illustrated in Fig. \ref{abs}{(d)}}. However, {this phenomenon is limited as the authors have observed that when} $\Nb$ is increased to {larger values}, the dips start to disappear {(though an illustration of this aspect is not present here)}. 
Although, the transmission behaviour, with a narrow transmission band, reported by 
\cite{shin2015} is different, {almost opposite in fact}, from the one reported in the present work ({as it is clear from} Fig. \ref{abs}), {what is common is that} some appropriate variations in the geometric arrangements of the cracks a favourable transmission/blocking of high frequency {waves} can be achieved.

\section{Conclusion}
{
The paper deals with the scattering due to a periodic array of staggered cracks in a two-dimensional lattice. Within the setting of finite cracks, the case of semi-infinite cracks is also considered as it is relevant when the length of cracks is much larger than the spacing in-between.
The latter can be seen as a discrete analogue of the work by Carlson and Heins \cite{HeinsI,HeinsII,HeinsIII}.
In this sense,  the present article considers a formulation on square lattice when there exists a discrete equivalent of an infinite array of parallel finite or semi-infinite rows with Neumann condition. }
Due to Floquet--Bloch theorem, the problem is reduced to that of the scattering due to a single crack on a lattice `waveguide' with Floquet boundary conditions on the outer rows. 
For the purpose of solving the problem for finite cracks, a discrete Green's function 
has been derived, which satisfies the Floquet periodic boundary conditions of the waveguide. The crack opening displacements and the exact solution 
of the scattering problem are obtained by the inversion of a Toeplitz matrix whose entries are found in terms of the Green's function \cite{sFK}. The limiting situation comprises of semi-infinite cracks which admits a refined solution obtained by the discrete Wiener--Hopf method. 
A low frequency approximation of the solution \cite{sK} in integral form recovers the classical continuum solution of Heins and Carlson; however, {such details are omitted but can be carried out following earlier work on continuum limit for singe crack \cite{sK,sFK,sConti}}. 

From a physical point of view, the transmission of the mechanical energy has been also been explored, via the notions of reflectance and transmittance \cite{brill,sWaveguide}, some results concerning which have been illustrated graphically. 
{A peculiar transmission behaviour is observed for certain range of incident wave and structural parameters wherein a narrow range of frequencies, in pass band of the infinite lattice, suffer almost complete reflection. Via an appropriate set of variations in the geometric arrangements of the cracks, therefore, the paper indicates the possibility of constructing some tunable atomic scale interfaces. The transmission of energy in such structures will have favourable {\em complete} transmission for most of the frequencies with the exception of an interesting small segment of high frequency waves that can be {\em blocked}}.
{In general,} the results presented in the article are expected to be useful in the study of scattering of elastic waves in crystalline materials as well as in the understanding of phonon transport at low temperature in systems involving superlattices.
The presence of periodically distributed interfaces in novel lattice structures {also} provides a rationale for the analysis of a simple case presented in the article. 

\section*{Acknowledgement}
GM acknowledges MHRD (India) and IITK for providing financial assistance in the form of Senior Research Fellowship.
BLS acknowledges the partial support of SERB MATRICS grant MTR/2017/000013.
This work has been available free of peer review on the arXiv since 12/2019.
The authors thank both anonymous reviewers for their constructive comments and suggestions. 

\printbibliography

\begin{appendix}

\section{`Waveguide' and wave modes}
\label{wdisperse}
\subsection{Wave modes in the `bulk'}
\label{appWmodeaway}
Consider the `wave modes' in right side of the strip ${\mathscr{S}}_0$, i.e., ${y}\in{\ZZ}_0^{{\mathtt{N}}-1}$.
By an application of the Floquet--Bloch condition \eqref{periodconditioncrackgen} for the equation of motion of the `upper' and `lower' boundary rows, i.e. 
${\su}_{{x}, {\mathtt{N}}}={\uppsi}{\su}_{{x}-{\mathtt{M}}, 0}, {\su}_{{x},-1}={\uppsi}^{-1}{\su}_{{x}+{\mathtt{M}}, {\mathtt{N}}-1}.$
With ${\su}_{{x}, {y}}({t})={{a}}_{{y}}{\exp}({-i{\upxi}{x}-i\la^{-1}{\upomega}{t}}), {y}\in{\ZZ}_{0}^{{\mathtt{N}}-1}$ and ${z}={\exp}({-i{\upxi}})$, it follows that \begin{eqn}
-{\upomega}^2{{a}}_{{y}}&=(1-{{\updelta}}_{{{y}}, {{\mathtt{N}}}-1}){{a}}_{{y}+1}+{{\updelta}}_{{{y}}, {{\mathtt{N}}}-1}{{a}}_{0}{\uppsi}{z}^{-{\mathtt{M}}}\\&
+(1-{{\updelta}}_{{{y}}, 0}){{a}}_{{y}-1}+{{\updelta}}_{{{y}}, 0}{{a}}_{{\mathtt{N}}-1}{\uppsi}^{-1}{z}^{{\mathtt{M}}}+2\cos{\upxi}{{a}}_{{y}}-4{{a}}_{{y}}, \quad\quad{y}\in{\ZZ}_{0}^{{\mathtt{N}}-1}.
\label{dHelmholtz_sq}
\end{eqn}
In particular, ${\upomega}$ is described by the general form 
${\upomega}^2=4-2\cos{\upxi}-2\cos{{\upeta}_{{\kappa}}}, {\kappa}\in{\ZZ}_1^{{\mathtt{N}}},$
where ${\upeta}_{{\kappa}}$ are determined by a specific condition, 
i.e.
$\sin({{\mathtt{N}}}+1){{\upeta}_{{\kappa}}}-\sin({\mathtt{N}}-1){{\upeta}_{{\kappa}}}-({\uppsi}{z}^{-{\mathtt{M}}}+{\uppsi}^{-1}{z}^{{\mathtt{M}}})\sin{\upeta}_{{\kappa}}=0,$
which can be expressed as
\begin{eqn}
{\mathtt{U}}_{{\mathtt{N}}}({\vartheta})-{\mathtt{U}}_{{\mathtt{N}}-2}({\vartheta})-({\uppsi}{z}^{-{\mathtt{M}}}+{\uppsi}^{-1}{z}^{{\mathtt{M}}})=0.
\label{eigvaleqn}
\end{eqn}
The eigenvectors ${{a}}_{({\kappa})}$ are given by (with ${\mathtt{C}}_{{\mathtt{N}}}$ normalization constant)
\begin{eqn}
{{a}}_{({\kappa}){y}}={\mathtt{C}}_{{\kappa};{\mathtt{N}}}(\sin({y}+1){{\upeta}_{{\kappa}}}+{\uppsi}^{-1}{z}^{{\mathtt{M}}}\sin({\mathtt{N}}-{y}-1){{\upeta}_{{\kappa}}}), \quad\quad{y}\in{\ZZ}_{0}^{{\mathtt{N}}-1},
\label{eigveceqn}
\end{eqn}
\begin{eqn}
\text{and }
{\mathtt{C}}^{-2}_{{\kappa};{\mathtt{N}}}&={\mathtt{N}}(1-\frac{1}{4}({\uppsi}{z}^{-{\mathtt{M}}}+{\uppsi}^{-1}{z}^{{\mathtt{M}}})^2).
\label{eigvecnorm}
\end{eqn}
Further, the branches of the dispersion relation are given by
\begin{eqn}
{\upomega}^2_{{\kappa}}=4\sin^2\frac{1}{2}{\upxi}+4\sin^2\frac{1}{2}{\upeta}_{{\kappa}}, {{\kappa}}\in{\ZZ}_{0}^{{\mathtt{N}}-1}. 
\end{eqn}
The group velocity is easily found to be given by (for ${{\kappa}}\in{\ZZ}_{0}^{{\mathtt{N}}-1}$)
\begin{eqn}
{{v}}_{{\kappa}}({\upxi})&=\pd{}{{\upxi}}{\upomega}_{{\kappa}}={\upomega}_{{\kappa}}^{-1}(\sin{\upxi}+\od{{\upeta}_{{\kappa}}}{{\upxi}}\sin{\upeta}_{{\kappa}})={\mathtt{N}}^{-1}{\upomega}_{{\kappa}}^{-1}({{\mathtt{N}}}\sin{\upxi}-{{\mathtt{M}}}{}\sin{\upeta}_{{\kappa}}).
\label{groupvelwaveguide}
\end{eqn}
Using $\sin n{{\upeta}_{{\kappa}}}/\sin{{\upeta}_{{\kappa}}}=({{{\lambda}}^{-n}-{{\lambda}}^n})/({{{\lambda}}^{-1}-{{\lambda}}})$, \eqref{eigveceqn} can be also written as
${a}_{\yy}
=C_{\NN}{\mathtt{U}}_{\NN-1}{{\lambda}}^{\NN-(\yy+1)}.$

\subsection{Wave modes between the cracks}
\label{appWmodeinside}
Consider the `wave modes' in left side of the strip ${\mathscr{S}}_0$, i.e., ${y}\in{\ZZ}_0^{{\mathtt{N}}-1}$.
With ${\su}_{{x}, {y}}({t})={{a}}_{({{\kappa}}){y}}{\exp}({-i{\upxi}{x}-i\la^{-1}{\upomega}{t}}), {y}\in{\ZZ}_{0}^{{\mathtt{N}}-1}$ and ${z}={\exp}({-i{\upxi}})$,
{the} wave modes are given by 
\begin{eqn}
{{a}}_{({{\kappa}}){{y}}}&={{a}}_{({{\kappa}})1}{\cos({{y}}+{\tfrac{1}{2}}){{\upeta}_{{\kappa}}}}/{\cos({\tfrac{1}{2}}{{\upeta}_{{\kappa}}})}, {{y}}\in{\ZZ}_0^{{\mathtt{N}}-1}, {{\upeta}_{{\kappa}}}={({{\kappa}}-1)\pi}/{{\mathtt{N}}}, {{\kappa}}\in{\ZZ}_1^{{\mathtt{N}}}. \label{oneDfree}
\end{eqn}
Note that ${{\kappa}}=1$ corresponds to ${{\upeta}_{{\kappa}}}_{1}=0$ for Neumann case, so that ${{a}}_{(1){{\nu}}}={{a}}_{(1)1}=1/\sqrt{{\mathtt{N}}}$, i.e. along the vertical direction it is a uniform translation of all ${{\mathtt{N}}}$ rows. 

\section{${{\mathcal{R}}}$ and ${{\mathcal{T}}}$ for finite cracks}
\label{RT}
\begin{figure}[h]
\centering
\includegraphics[width=.75\textwidth]{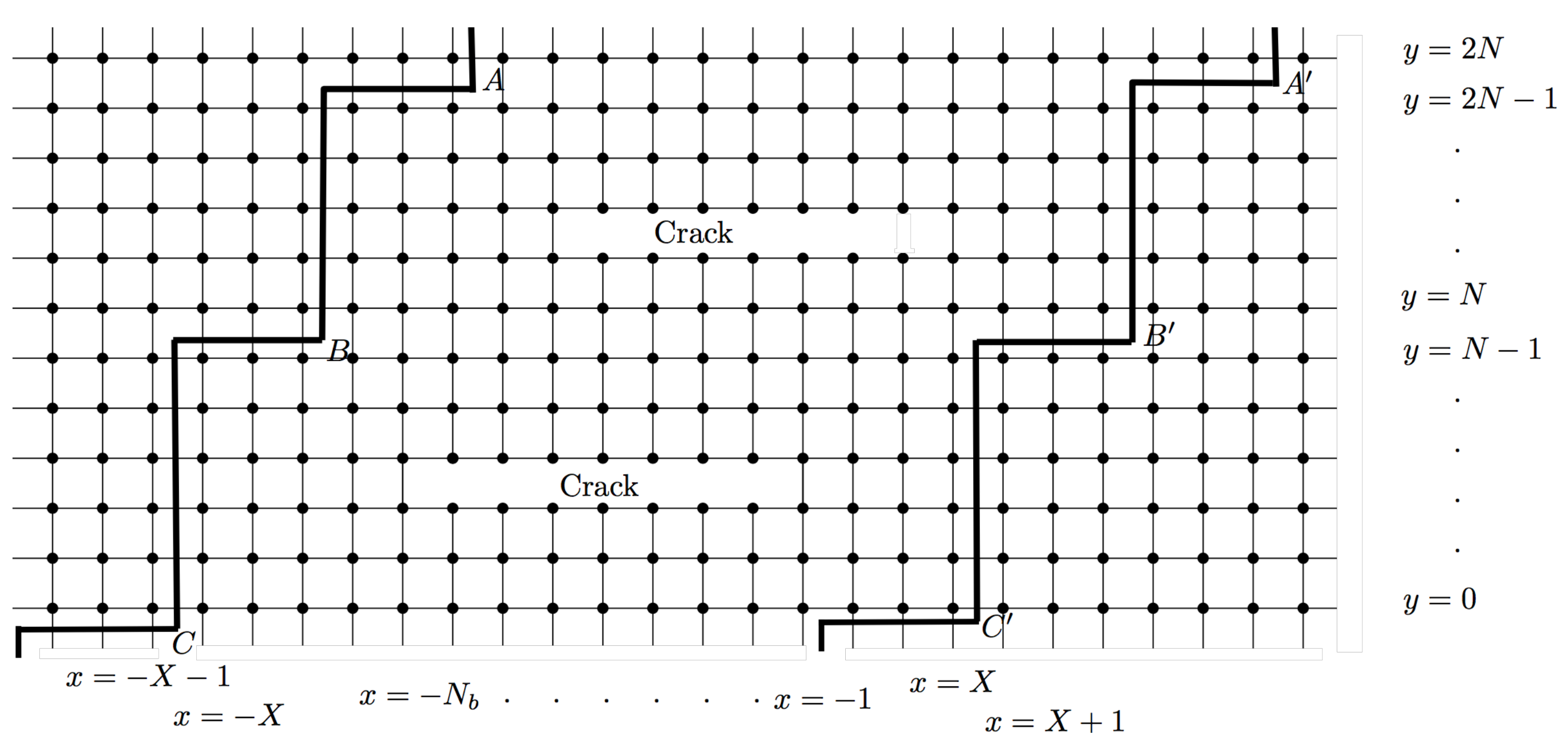}
\caption{Schematic of the square lattice with the array of staggered finite cracks for the calculation of the energy flux in the incident, reflected and transmitted waves across the boundaries B'C' and BC.}
\label{bound}
\vspace{-7mm}
\end{figure}
The reflectance ${{\mathcal{R}}}$ (resp. transmittance ${\mathcal{T}}$) is the ratio of the energy flux in the outgoing wave ahead (resp. behind) of the cracks to the energy flux carried by the incident wave \cite{brill} (across a boundary shown in Fig. \ref{bound} with thick solid lines). 
Since the cracks in the array are staggered with respect to each other, the energy flux is calculated across a boundary shown in Fig. \ref{bound} with thick solid lines. In Fig. \ref{bound}, $X$ is taken far away from the cracks. Along the boundary BC (resp., B'C'), there are $\MM$ vertical bonds and $\NN$ horizontal bonds that are broken. 
Following \cite{brill}, the energy flux carried by the incident wave across the boundary B'C' (Fig. \ref{bound}) is found to be 
\begin{equation}
\begin{aligned}
\mathtt{W}^i&=\Re\sum\nolimits_{n=0}^{\NN-1}((\uu_{X+1,n}^i-\uu_{X,n}^i )\overline{(-i \omega \uu_{X,n}^i)})\\&+\Re\sum\nolimits_{m=1}^{\MM}((\uu_{X+m,\NN-1}^i-{{\uppsi}}\uu_{X+m-\MM,0}^i)\overline{(-i\omega {{\uppsi}}\uu_{X+m-\MM,0}^i)},
\label{inputwork}
\end{aligned}
\end{equation}
and the energy flux in the outgoing wave ahead of the cracks, that is, the energy flux in the reflected wave across the boundary B'C' (as in Fig. \ref{bound}) can be written as 
\begin{equation}
\begin{aligned}
\mathtt{W}^r&=\Re\sum\nolimits_{n=0}^{\NN-1}((\uu_{X,n}-\uu_{X+1,n})\overline{(-i\omega \uu_{X+1,n})})\\&+\Re\sum\nolimits_{m=1}^{\MM}(({{\uppsi}}\uu_{X+m-\MM,0}-\uu_{X+m,\NN-1})\overline{(-i\omega \uu_{X+m,\NN-1})}.
\label{reflectedwork}
\end{aligned}
\end{equation}
Similarly, the energy flux in the outgoing wave behind the cracks, that is, the energy flux carried by the transmitted wave across the boundary BC (see Fig. \ref{bound}) is determined as 
\begin{equation}
\begin{aligned}
\mathtt{W}^t&=\Re\sum\nolimits_{n=0}^{\NN-1}((\uu_{-X,n}^t-\uu_{-X-1,n}^t)\overline{(-i\omega \uu_{-X-1,n}^t)}\\&+\Re\sum\nolimits_{m=1}^{\MM}((\uu_{-X-1-m,\NN-1}^t-{{\uppsi}}\uu_{-X-1+m-\MM,0}^t)\overline{(-i\omega {{\uppsi}}\uu_{-X-1+m-\MM,0}^t)}.
\label{transmittedwork}
\end{aligned}
\end{equation}
Using \eqref{inputwork}, \eqref{reflectedwork}, 
and \eqref{transmittedwork}, the reflectance and transmittance can be written as
\begin{eqnarray}
{{\mathcal{R}}}={\mathtt{W}^r}/{\mathtt{W}^i},\label{reflectance} 
{\mathcal{T}}={\mathtt{W}^t}/{\mathtt{W}^i}, 
\end{eqnarray}
respectively.
For $\yy>\yy_0,$ in the numerator of \eqref{Gyfinal},
\begin{equation}
\begin{aligned}
&{\mathtt{U}}_{\NN-|\yy-\yy_0|-1}
+{\mathtt{U}}_{|\yy_0-\yy|-1}
({\uppsi}\varz^{-\MM})^{\text{sign}(\yy-\yy_0)}\\
&=({{{\lambda}}^{-1}-{{\lambda}}})^{-1}({{\lambda}}^{-\NN+|\yy-\yy_0|}-{{\lambda}}^{\NN-|\yy-\yy_0|}+{{\lambda}}^{-\NN\text{sign}(\yy-\yy_0)}({{\lambda}}^{-|\yy-\yy_0|}-{{\lambda}}^{|\yy-\yy_0|}))\\
&=({{{\lambda}}^{-1}-{{\lambda}}})^{-1}({{\lambda}}^{-\NN}-{{\lambda}}^{\NN}){{\lambda}}^{-(\yy-\yy_0)},
\end{aligned}
\end{equation}
while, analogously, for $\yy<\yy_0,$ in the numerator of \eqref{Gyfinal},
\begin{equation}
\begin{aligned}
&{\mathtt{U}}_{\NN-|\yy-\yy_0|-1}
+{\mathtt{U}}_{|\yy_0-\yy|-1}
({\uppsi}\varz^{-\MM})^{\text{sign}(\yy-\yy_0)}
=({{{\lambda}}^{-1}-{{\lambda}}})^{-1}({{\lambda}}^{-\NN}-{{\lambda}}^{\NN}){{\lambda}}^{-(\yy-\yy_0)}.
\end{aligned}
\end{equation}
Hence, for $\yy\in{\ZZ}_{0}^{\NN-1}$,
{${\mathtt{U}}_{\NN-|\yy-\yy_0|-1}
+{\mathtt{U}}_{|\yy_0-\yy|-1}
({\uppsi}\varz^{-\MM})^{\text{sign}(\yy-\yy_0)}
={a}_{\yy}{{{\lambda}}^{-\NN+\yy_0+1}}/{C_{\NN}}.$}
In the context of the expression {\eqref{Gyfinal} of Green's function ${\GG}_{\yy}^{{\mathrm{F}}}$}, let
\begin{equation}
\begin{aligned}
D={2{\mathtt{T}}_\NN
-({\uppsi}\varz^{-\MM}+{\uppsi}^{-1}\varz^{\MM})}=D_+D_-,
\end{aligned}
\label{Ddef}
\end{equation}
where $D_-$ (resp. $D_+$) contains those zeros of $D$ which lie outside (resp. inside) the unit circle for $\oo_2>0$ \eqref{damping}. Recall that $\oo_2\to0+$.
Thus, for $\xx\to-\infty$, it follows from \eqref{fourierinvG} and \eqref{Gyfinal}, by an application of the complex residue calculus and \eqref{Ddef}, that,
\begin{equation}
\begin{aligned}
\GG_{\xx,\yy}&\sim-\sum\nolimits_{D_-=0}\frac{\varz^{-\xx_0-1}}{D'(\varz)}\frac{{{\lambda}}^{-\NN+\yy_0+1}}{C_{\NN}}{a}_{\yy}\varz^{\xx}
=-\sum\nolimits_{D_-=0}\frac{\varz^{-\xx_0-1}}{D'(\varz)}\frac{{{\lambda}}^{-\NN}-{{\lambda}}^{\NN}}{{{\lambda}}^{-1}-{{\lambda}}}{{\lambda}}^{\yy_0}\varz^{\xx}{{\lambda}}^{-\yy}.
\end{aligned}
\end{equation}
Similarly, 
for $\xx\to+\infty$,
$\GG_{\xx,\yy}\sim
\sum\nolimits_{D_+=0}\frac{\varz^{-\xx_0-1}}{D'(\varz)}\frac{{{\lambda}}^{-\NN}-{{\lambda}}^{\NN}}{{{\lambda}}^{-1}-{{\lambda}}}{{\lambda}}^{\yy_0}\varz^{\xx}{{\lambda}}^{-\yy}.$
Let
\begin{equation}
\begin{aligned}
f(\varz)=\sum\nolimits_{l=-\Nb}^{-1}(\mathtt{v}_l+\mathtt{v}_l^i)\varz^{-l}.
\end{aligned}
\end{equation}
Thus, for $\xx\to-\infty$,
\begin{equation}
\begin{aligned}
\uu_{\xx,\yy}
&\sim\sum\nolimits_{D_+=0}\frac{\varz^{-1}}{D'(\varz)}\frac{{{\lambda}}^{-\NN}-{{\lambda}}^{\NN}}{{{\lambda}}^{-1}-{{\lambda}}}{{\lambda}}^{\Na}\varz^{\xx}{{\lambda}}^{-\yy}(1-{{\lambda}}^{-1})\sum\nolimits_{l=-\Nb}^{-1}(\mathtt{v}_l+\mathtt{v}_l^i)\varz^{-l}\\
&\sim\sum\nolimits_{D_+=0}(\frac{(1-{{\lambda}}^{-1})\varz^{-1}}{D'(\varz)}{\mathtt{U}}_{\NN-1}{{\lambda}}^{\Na})\varz^{\xx}{{\lambda}}^{-\yy}f(\varz)\\
&\sim\sum\nolimits_{D_+=0}K_{\varz}\varz^{\xx}{a}_{\yy}, K_{\varz}=\frac{(1-{{\lambda}}^{-1})\varz^{-1}}{D'(\varz)}\frac{{{\lambda}}^{-\NN+\Na+1}}{C_{\NN}}f(\varz).
\end{aligned}
\end{equation}
Similarly,
for $\xx\to+\infty$,
$\uu_{\xx,\yy}\sim-\sum\nolimits_{D_-=0}K_{\varz}\varz^{\xx}{a}_{\yy}.$  
The incident energy flux in a specific mode is given by $-({1}/{2})|\tilde{K}|^2\oo^2\mathcal{V}^{i}$ where $-\mathcal{V}^{i}$ is the group velocity inside the `waveguide'.
Hence, the reflectance and transmittance of the scatterer array are given by
\begin{equation}
\begin{aligned}
{{\mathcal{R}}}&=
\frac{\sum{E}^{\tilde{r}}_{\text{reflected}}}{{E}^{\tilde{r}}_{\text{incident}}}=\frac{\sum\nolimits_{D_+=0}|K_{\varz}|^2\mathcal{V}}{-|\tilde{K}|^2\mathcal{V}^{i}}, 
{\mathcal{T}}&=
\frac{\sum{E}^{\tilde{r}}_{\text{transmitted}}}{{E}^{\tilde{r}}_{\text{incident}}}=\frac{\sum\nolimits_{D_+=0}|K_{\varz}|^2\mathcal{V}}{-|\tilde{K}|^2\mathcal{V}^{i}},
\end{aligned}
\end{equation}
respectively. 
{In this} expression of the reflectance ${{\mathcal{R}}}$,
\begin{equation}
\begin{aligned}
\frac{|K_{\varz}|^2\mathcal{V}}{-|\tilde{K}|^2\mathcal{V}^{i}}&=-\frac{\mathcal{V}}{\mathcal{V}^{i}}|\frac{(1-{{\lambda}}^{-1})\varz^{-1}}{D'(\varz)}|^2|\frac{{{\lambda}}^{-\NN+\Na+1}}{C_{\NN}}|^2|f(\varz)|^2\\
&=-\frac{i\varz}{-2\oo\NN {\mathtt{U}}_{\NN-1}
}\frac{1}{\mathcal{V}^{i}}\frac{(1-{{\lambda}}^{-1})\overline{(1-{{\lambda}}^{-1})}}{\overline{D'(\varz)}}\NN(1-\frac{1}{4}({\uppsi}\varz^{-\MM}+{\uppsi}^{-1}\varz^{\MM})^2)|f(\varz)|^2\\
&=\frac{i\varz}{-2\oo {\mathtt{U}}_{\NN-1}
}\frac{1}{\mathcal{V}^{i}}\frac{{\HH}}{\overline{D_-(\varz)}\overline{D'_+(\varz)}}(1-\frac{1}{4}({\uppsi}\varz^{-\MM}+{\uppsi}^{-1}\varz^{\MM})^2)|f(\varz)|^2.
\end{aligned}
\end{equation}

\section{${{\mathcal{R}}}$ and ${{\mathcal{T}}}$
for semi-finite cracks: bulk incidence}
\label{RTsemi}

For the purpose of the manipulations presented below, consider ${\vartheta}({z}, {\upomega})$, i.e. as a function of ${z}$ and ${\upomega}$; same consideration applies to other relevant functions. 
Thus, the following relations are obtained concerning the group velocity of wave modes, on either side of the scatterer,
\begin{eqn}
\frac{{{v}}}{{\mathscr{N}}'({z})}=\frac{i{z}}{\pd{}{{\upomega}}{\mathscr{N}}({z}, {\upomega})}|_{{{z}}={\exp}({-i{\upxi}})}, \quad
\frac{{{v}}}{{\mathscr{D}}'({z})}=\frac{i{z}}{\pd{}{{\upomega}}{\mathscr{D}}({z}, {\upomega})}|_{{{z}}={\exp}({-i{\upxi}})}.
\label{NDLder}
\end{eqn} 

Note that by reference to \eqref{Lk_array}, {we have} ${\mathscr{N}}({{z}})={{\HH}}({z}){\mathtt{U}}_{{\mathtt{N}}-1}({\vartheta}), {\mathscr{D}}({{z}})=2{\mathtt{T}}_{{\mathtt{N}}}({\vartheta})-({z}^{{\mathtt{M}}}{z}_{{P}}^{-{\mathtt{M}}}{\lambda}_{{P}}^{{\mathtt{N}}}+{z}^{-{\mathtt{M}}}{z}_{{P}}^{{\mathtt{M}}}{\lambda}_{{P}}^{-{\mathtt{N}}}),$ where ${\vartheta}=\frac{1}{2}{\mathpzc{Q}}({{z}}), {\mathpzc{Q}}={{\HH}}+2.$ Invoking \eqref{NDLder}${}_1$ (while using 
${\mathtt{U}}'_{n}={(n+1){\mathtt{T}}_{n+1}}/({{\vartheta}^2-1})$ when ${\mathtt{U}}_{n}$ is zero)
with ${z}$ such that ${\mathscr{N}}({z})=0$, {(let $\RR({z}):=\HH({z})+4$, recall \eqref{HHdef})}
\begin{eqn}
\pd{}{{\upomega}}{\mathscr{N}}({z}, {\upomega})&={{\HH}}({z}){\mathtt{N}}\frac{2(\pd{}{{\upomega}}{\mathpzc{Q}}({z}))}{{\mathpzc{Q}}({z})^2-4}{\mathtt{T}}_{{\mathtt{N}}}({\vartheta})\text{ or }(\pd{}{{\upomega}}{{\HH}}({z})){\mathtt{U}}_{{\mathtt{N}}-1}({\vartheta})\\&
=-4{\upomega}\frac{{{\HH}}({z}){\mathtt{N}}{\mathtt{T}}_{{\mathtt{N}}}({\vartheta})}{{\mathpzc{Q}}({z})^2-4}\text{ or }-2{\upomega}{\mathtt{U}}_{{\mathtt{N}}-1}({\vartheta})=-4{\upomega}{\mathtt{N}}\frac{{\mathtt{T}}_{{\mathtt{N}}}({\vartheta})}{{\RR}({z})}\text{ or }-2{\upomega}{\mathtt{U}}_{{\mathtt{N}}-1}({\vartheta}).
\label{Nweqk}
\end{eqn} 
Similarly, invoking \eqref{NDLder}${}_2$, while (using 
${\mathtt{T}}'_n=n{\mathtt{U}}_{n-1}$)
with ${z}$ such that ${\mathscr{D}}({z})=0$, 
{$\pd{}{{\upomega}}{\mathscr{D}}({z}, {\upomega})={\mathtt{N}}(\pd{}{{\upomega}}{\mathpzc{Q}}){\mathtt{U}}_{{\mathtt{N}}-1}({\vartheta})
=-2{\upomega}{\mathtt{N}}{\mathtt{U}}_{{\mathtt{N}}-1}({\vartheta}).$}
Hence, behind and ahead of the scatterer, respectively,
{
${{v}}/{\mathscr{N}}'({z})=i{z}{\RR}/(-4{\upomega}{\mathtt{N}}{\mathtt{T}}_{{\mathtt{N}}}({\vartheta}))$ or $i{z}/(-2{\upomega}{\mathtt{U}}_{{\mathtt{N}}-1}({\vartheta})),$
${{v}}/{\mathscr{D}}'({z})=i{z}/(-2{\upomega}{\mathtt{N}}{\mathtt{U}}_{{\mathtt{N}}-1}({\vartheta})).$}
Using the normal modes for a lattice strip of width ${\mathtt{N}}$, which is free on upper and lower boundary, (using $\sin{\mathtt{N}}{\upeta}_{{\kappa}}=0$, which implies $\cos({\mathtt{N}}-\frac{1}{2}){\upeta}_{{\kappa}}=\cos{\mathtt{N}}{\upeta}_{{\kappa}}\cos\frac{1}{2}{\upeta}_{{\kappa}}$)
{$|{{a}}_{-({{\kappa}}_{{z}})0}-{\uppsi}^{-1}{z}^{{\mathtt{M}}}{{a}}_{-({{\kappa}}_{{z}}){\mathtt{N}}-1}|^2=\frac{1}{2{\mathtt{N}}}\cos{\mathtt{N}}{{\upeta}_{{\kappa}}}{\RR}({z}){\mathscr{D}}({{z}}).$}
But ${\mathtt{U}}_{{\mathtt{N}}-1}=0$ implies that ${\mathtt{U}}_{{\mathtt{N}}-2}=-{\mathtt{U}}_{{\mathtt{N}}},$ i.e., $2{\mathtt{T}}_{{\mathtt{N}}}=2{\mathtt{U}}_{{\mathtt{N}}}.$ More directly, ${\vartheta}=\cos\frac{j\pi}{{\mathtt{N}}},$ so that $2\cos^2\frac{1}{2}{{\upeta}_{{\kappa}}}=1+\cos{\upeta}_{{\kappa}}=1+{\vartheta}=2\cos^2\frac{j\pi}{2{\mathtt{N}}}, {\upeta}_{{\kappa}}=\pm\frac{j\pi}{{\mathtt{N}}}$. Also
$({\vartheta}-1){\mathtt{U}}_{{\mathtt{N}}}=\frac{1}{2}({\mathtt{U}}_{{\mathtt{N}}+1}+{\mathtt{U}}_{{\mathtt{N}}-1}-2{\mathtt{U}}_{{\mathtt{N}}})$. Note that
${{\mathtt{U}}}_{{\mathtt{N}}}^2-{{\mathtt{U}}}_{{\mathtt{N}}-1}{{\mathtt{U}}}_{{\mathtt{N}}+1}=1$ which implies ${{\mathtt{U}}}_{{\mathtt{N}}}^2=1$ when ${{\mathtt{U}}}_{{\mathtt{N}}-1}=0,$ so that ${{\mathtt{T}}}_{{\mathtt{N}}}={{\mathtt{U}}}_{{\mathtt{N}}}=\pm1.$
The transmittance is given by
\begin{eqn}
{{\mathcal{T}}}&=\frac{({\mathtt{N}}{{v}}({\upxi}_{{P}}))^{-1}}{|{{\mathpzc{L}_{{\mathtt{N}}}}}_{{}+}({{z}}_{{P}})|^2}\sum\nolimits_{{{z}}\in{\mathcal{Z}^-}}\frac{|-2i\sin\frac{1}{2}{{\upeta}_{{\kappa}}}({\vartheta}_{{\mathtt{N}}}({{z}}_{{P}}))|^2}{\frac{1}{2{\mathtt{N}}}\cos{\mathtt{N}}{{\upeta}_{{\kappa}}}{\RR}({z}){\mathscr{D}}({{z}})}\frac{{\mathscr{D}}({{z}})}{\overline{{\mathscr{N}}'({{z}})}}({{v}}({\upxi}))\frac{\overline{{\mathscr{D}}_-({{z}}){\mathscr{N}}_+({{z}})}}{{\mathscr{N}}'_-({{z}}){\mathscr{D}}_+({{z}})}\frac{1}{\overline{{z}}-\overline{{z}}_{{P}}}\frac{1}{{{z}}-{{z}}_{{P}}} \\&=\frac{({\mathtt{N}}{{v}}({\upxi}_{{P}}))^{-1}}{|{{\mathpzc{L}_{{\mathtt{N}}}}}_{{}+}({{z}}_{{P}})|^2}\sum\nolimits_{{{z}}\in{\mathcal{Z}^-}}\frac{4\sin^2\frac{1}{2}{{\upeta}_{{\kappa}}}({\vartheta}_{{\mathtt{N}}}({{z}}_{{P}}))}{\frac{1}{2{\mathtt{N}}}\cos{\mathtt{N}}{{\upeta}_{{\kappa}}}{\RR}({z}){\mathscr{D}}({{z}})}\frac{-i\overline{{z}}{\mathscr{D}}({{z}})}{\overline{{\mathtt{N}}\frac{2(-2{\upomega})}{{\RR}}{{\mathtt{T}}}_{{\mathtt{N}}}({\vartheta})}}\frac{\overline{{\mathscr{D}}_-({{z}}){\mathscr{N}}_+({{z}})}}{{\mathscr{N}}'_-({{z}}){{\mathscr{D}}_+({{z}})}}\frac{1}{\overline{{z}}-\overline{{z}}_{{P}}}\frac{1}{{{z}}-{{z}}_{{P}}}\\&=-2i\frac{({\mathtt{N}}{{v}}({\upxi}_{{P}}))^{-1}}{|{{\mathpzc{L}_{{\mathtt{N}}}}}_{{}+}({{z}}_{{P}})|^2}\sum\nolimits_{{{z}}\in{\mathcal{Z}^-}}\frac{{{\HH}}({{z}}_{{P}})}{{{\mathtt{T}}}_{{\mathtt{N}}}({\vartheta})}\frac{1}{\overline{4{\upomega}{{\mathtt{T}}}_{{\mathtt{N}}}({\vartheta})}}\frac{\overline{{\mathscr{D}}_-({{z}}){\mathscr{N}}_+({{z}})}}{{\mathscr{N}}'_-({{z}}){{\mathscr{D}}_+({{z}})}}\frac{\overline{{z}}}{\overline{{z}}-\overline{{z}}_{{P}}}\frac{1}{{{z}}-{{z}}_{{P}}}\\
&=\frac{1}{2}{i}\frac{{\upomega}^{-1}{{\HH}}({{z}}_{{P}})}{({{v}}({\upxi}_{{P}}))|{{\mathpzc{L}_{{\mathtt{N}}}}}_{{}+}({{z}}_{{P}})|^2{\mathtt{N}}}\sum\nolimits_{{{z}}\in{\mathcal{Z}^-}}\frac{\overline{{\mathscr{D}}_-({{z}}){\mathscr{N}}_+({{z}})}}{{\mathscr{N}}'_-({{z}}){\mathscr{D}}_+({{z}})}\frac{{{z}}_{{P}}}{({{z}}-{{z}}_{{P}})^2}.
\label{Trans_k_sq_P}
\end{eqn}

Using \eqref{eigvaleqn}, \eqref{eigveceqn}, and \eqref{eigvecnorm} (using $2{\mathtt{T}}_{{\mathtt{N}}}({\vartheta})=({\uppsi}^{-1}{z}^{{\mathtt{M}}}+{\uppsi}{z}^{-{\mathtt{M}}})=({z}^{{\mathtt{M}}}{z}_{{P}}^{-{\mathtt{M}}}{\lambda}_{{P}}^{{\mathtt{N}}}+{z}^{-{\mathtt{M}}}{z}_{{P}}^{{\mathtt{M}}}{\lambda}_{{P}}^{-{\mathtt{N}}})$), 
$|{{a}}_{+({{\kappa}}_{{z}})0}-{\uppsi}^{-1}{z}^{{\mathtt{M}}}{{a}}_{+({{\kappa}}_{{z}}){\mathtt{N}}-1}|^2=-\frac{{\mathscr{N}}({z})}{{\mathtt{N}}{\mathtt{U}}_{{\mathtt{N}}-1}({\vartheta})}.$
{In} \eqref{vm0asym}, 
${\mathscr{D}}'({{z}})=2{\vartheta}'({{z}}){{\mathtt{T}}}'_{{\mathtt{N}}}({\vartheta})={{\HH}}'({{z}}){\mathtt{N}}{\mathtt{U}}_{{\mathtt{N}}-1}({\vartheta}),$ 
${\mathscr{N}}'({{z}})={{\HH}}({{z}}){\vartheta}'({{z}}){{\mathtt{U}}}'_{{\mathtt{N}}-1}({\vartheta})
={{\HH}}({{z}}){\vartheta}'({{z}}){{\mathtt{U}}}'_{{\mathtt{N}}-1}({\vartheta})={{\HH}}({{z}}){\mathtt{N}}\frac{2{{\HH}}'}{{{\HH}}{\RR}}{{\mathtt{T}}}_{{\mathtt{N}}}({\vartheta})={\mathtt{N}}\frac{2{{\HH}}'}{{\RR}}{{\mathtt{T}}}_{{\mathtt{N}}}({\vartheta})$ or ${{\HH}}'({{z}}){{\mathtt{U}}}_{{\mathtt{N}}-1}({\vartheta}),$
where ${\vartheta}=\cos\theta, \theta\in[0, \pi]$. 
The ratio of the total mechanical energy flow (through all outgoing lattice waves) to the right and rate of the total mechanical energy influx (through the incident wave) can be expressed as
\begin{eqn}
{{\mathcal{R}}}&=\frac{({\mathtt{N}}{{v}}({\upxi}_{{P}}))^{-1}}{|{{\mathpzc{L}_{{\mathtt{N}}}}}_{{}+}({{z}}_{{P}})|^2}\sum\nolimits_{{{z}}\in{\mathcal{Z}^+}}\frac{|-2i\sin\frac{1}{2}{{\upeta}_{{\kappa}}}({\vartheta}_{{\mathtt{N}}}({{z}}_{{P}}))|^2}{-\frac{{\mathscr{N}}({z})}{{\mathtt{N}}{\mathtt{U}}_{{\mathtt{N}}-1}({\vartheta})}}\frac{1}{\overline{{z}}-\overline{{z}}_{{P}}} \frac{\overline{{\mathscr{D}}_-({{z}})}\overline{{\mathscr{N}}_+({{z}})}}{\overline{{\mathscr{D}}'({{z}})}}\frac{1}{{{z}}-{{z}}_{{P}}} \frac{{\mathscr{N}}({{z}})}{{\mathscr{D}}'_+({{z}}){\mathscr{N}}_-({{z}})}|{{v}}({\upxi})|\\
&=-\frac{({\mathtt{N}}{{v}}({\upxi}_{{P}}))^{-1}}{|{{\mathpzc{L}_{{\mathtt{N}}}}}_{{}+}({{z}}_{{P}})|^2}\sum\nolimits_{{{z}}\in{\mathcal{Z}^+}}\frac{{{{\mathtt{N}}{\mathtt{U}}_{{\mathtt{N}}-1}({\vartheta})}|-2i\sin\frac{1}{2}{{\upeta}_{{\kappa}}}({\vartheta}_{{\mathtt{N}}}({{z}}_{{P}}))|^2}}{(\overline{{z}}-\overline{{z}}_{{P}})({{z}}-{{z}}_{{P}})} \frac{\overline{{\mathscr{D}}_-({{z}})}\overline{{\mathscr{N}}_+({{z}})}}{{\mathscr{D}}'_+({{z}}){\mathscr{N}}_-({{z}})}\frac{-i\overline{{z}}}{-2{\upomega}{\mathtt{N}}{\mathtt{U}}_{{\mathtt{N}}-1}({\vartheta})}\\
&=-\frac{({\mathtt{N}}{{v}}({\upxi}_{{P}}))^{-1}}{|{{\mathpzc{L}_{{\mathtt{N}}}}}_{{}+}({{z}}_{{P}})|^2}\sum\nolimits_{{{z}}\in{\mathcal{Z}^+}}{{\HH}}({z}_{{P}})\frac{1}{\overline{{z}}-\overline{{z}}_{{P}}}\frac{1}{{{z}}-{{z}}_{{P}}} \frac{\overline{{\mathscr{D}}_-({{z}})}\overline{{\mathscr{N}}_+({{z}})}}{{\mathscr{D}}'_+({{z}}){\mathscr{N}}_-({{z}})}\frac{-i\overline{{z}}}{-2{\upomega}}\\&=-\frac{1}{2}{i}\frac{{\upomega}^{-1}{{\HH}}({{z}}_{{P}})}{({{v}}({\upxi}_{{P}}))|{{\mathpzc{L}_{{\mathtt{N}}}}}_{{}+}({{z}}_{{P}})|^2{\mathtt{N}}}\sum\nolimits_{{{z}}\in{\mathcal{Z}^+}}\frac{\overline{{\mathscr{D}}_-({{z}}){\mathscr{N}}_+({{z}})}}{{\mathscr{D}}'_+({{z}}){\mathscr{N}}_-({{z}})}\frac{\overline{{z}}}{({{z}}-{{z}}_{{P}})(\overline{{z}}-\overline{{z}}_{{P}})} \\
&=\frac{1}{2}{i}\frac{{\upomega}^{-1}{{\HH}}({{z}}_{{P}})}{({{v}}({\upxi}_{{P}}))|{{\mathpzc{L}_{{\mathtt{N}}}}}_{{}+}({{z}}_{{P}})|^2{\mathtt{N}}}\sum\nolimits_{{{z}}\in{\mathcal{Z}^+}}\frac{\overline{{\mathscr{D}}_-({{z}}){\mathscr{N}}_+({{z}})}}{{\mathscr{D}}'_+({{z}}){\mathscr{N}}_-({{z}})}\frac{{{z}}_{{P}}}{({{z}}-{{z}}_{{P}})^2}.
\label{Ref_k_sq_P}
\end{eqn}
For the incident wave from the bulk lattice, since ${\mathscr{D}}_-({{z}}_{{P}})=0$, ${{v}}({\upxi}_{{P}})$ is given by
${{{v}}({\upxi}_{{P}})}={i{z}_{{P}}{{\mathscr{D}}_+({{z}}_{{P}})}{{\mathscr{D}}'_-({{z}}_{{P}})}}/({-2{\upomega}{\mathtt{N}}{\mathtt{U}}_{{\mathtt{N}}-1}({\vartheta}({{z}}_{{P}}))}).$
The coefficient in front of the sum can be simplified as
\begin{eqn}
\frac{1}{2}{\upomega}^{-1}{i}\frac{{{\HH}}({{z}}_{{P}})}{({{v}}({\upxi}_{{P}}))|{{\mathpzc{L}_{{\mathtt{N}}}}}_{{}+}({{z}}_{{P}})|^2{\mathtt{N}}}&=\frac{{{z}}_{{P}}{\mathscr{N}}_-({{z}}_{{P}}){\mathscr{D}}_+({{z}}_{{P}})}{\overline{{\mathscr{D}}'_-({{z}}_{{P}})}\overline{{\mathscr{N}}_+({{z}}_{{P}})}}.
\label{CRexp}
\end{eqn}

\end{appendix}

\end{document}